\documentclass[12pt,a4paper]{article}
\pdfoutput=1
\usepackage{graphicx}
\usepackage{subfigure}
\usepackage{threeparttable}
\usepackage{array}
\usepackage{multirow}
\usepackage{amsmath}
\usepackage{hyperref}
\hypersetup{pdfauthor={Ruifeng Yuan, Chengwen Zhong},pdftitle={An immersed-boundary method for compressible viscous flow and its application in gas-kinetic BGK scheme}}

\begin{document}


\title{An immersed-boundary method for compressible viscous flow and its application in gas-kinetic BGK scheme}
\renewcommand{\thefootnote}{\fnsymbol{footnote}}
\author{Ruifeng Yuan\footnotemark[1], Chengwen Zhong\footnotemark[1]}
\footnotetext[1]{National Key Laboratory of Science and Technology on Aerodynamic Design and Research, Northwestern Polytechnical University, Xi'an, Shaanxi 710072, China}
\footnotetext{\emph{Email addresses:} xyrfx@mail.nwpu.edu.cn (Ruifeng Yuan), zhongcw@nwpu.edu.cn (Chengwen Zhong)}
\date{Oct. 1, 2016}
\maketitle


\rule[-5pt]{\textwidth}{0.5pt}
\begin{abstract}
An immersed-boundary (IB) method is proposed and applied in the gas-kinetic BGK scheme to simulate incompressible/compressible viscous flow with stationary/moving boundary. In present method the ghost-cell technique is adopted to fulfill the boundary condition on the immersed boundary. A novel idea ``local boundary determination'' is put forward to identify the ghost cells, each of which may have several different ghost-cell constructions corresponding to different boundary segments, thus eliminating the singularity of the ghost cell. Furthermore, the so-called ``fresh-cell'' problem when implementing the IB method in moving-boundary simulation is resolved by a simple extrapolation in time. The method is firstly applied in the gas-kinetic BGK scheme to simulate the Taylor-Couette flow, where the second-order spatial accuracy of the method is validated and the ``super-convergence'' of the BGK scheme is observed. Then the test cases of supersonic flow around a stationary cylinder, incompressible flow around an oscillating cylinder and compressible flow around a moving airfoil are conducted to verify the capability of the present method in simulating compressible flows and handling the moving boundary.
~\\

\end{abstract}
\rule[5pt]{\textwidth}{0.5pt}


\section{Introduction}

In recent years, rising attention has been paid to a class of non-boundary-conforming grid methods called the immersed-boundary (IB) method for its convenience in handling problems with complex as well as moving boundaries. In the IB method, the grid points don't have to coincide with the solid boundary and a simple uniform or nonuniform Cartesian grid can be used, which will make the grid generation much easier than the conventional boundary-conforming grid method and guarantee the grid quality for even an extremely complex boundary. Without grid transformation and regeneration, the IB method is very applicable to moving boundary problems with relatively simple solution procedure and significantly lower computational cost compared to the conventional boundary-conforming grid method, especially for the problem involving multi-body movement, such as the problem concerns the interaction between fluid and fragments of aircraft, where the laborious grid regeneration or transformation can drive one crazy.

According to Mittal and Iaccarino \cite{mittal2005immersed}, most of the existing IB methods can be categorized into two groups, i.e. the continuous forcing approach and the direct (or discrete) forcing approach. In the continuous forcing approach, a forcing term is added to the continuous governing equations before they are discretized. The force is determined from the boundary condition on the Lagrangian boundary point and the discrete Dirac delta is usually used to link the Lagrangian and the Eulerian variables. The major advantage for the continuous forcing approach is that it can be easily implemented in an existing numerical scheme to deal with problems with complex or moving boundaries. The main drawback is the low spatial accuracy (generally first-order local spatial accuracy) due to the use of the discrete delta function. Besides, this class of IB methods is rarely used in the compressible problem.

In the direct forcing approach, the boundary condition is imposed by directly constructing variables on the grid points or cells near the immersed boundary involved in the calculation stencil of the numerical scheme. In contrast to the continuous forcing approach, for this category of IB methods the implementation is more complex and the inclusion of boundary motion is more difficult. But the advantage is that it allows a sharp representation of the immersed boundary and second-order local spatial accuracy can be obtained for some of the approaches. The pioneer work for this category of methods was done by Mohd-Yusof \cite{mohd1997combined} and Fadlun et al.~\cite{fadlun2000combined}. De Palma et al.~\cite{de2006immersed}, Ghias et al.~\cite{ghias2007sharp} and de Tullio et al.~\cite{de2007immersed} extended the direct forcing approach to the viscous compressible problem. Among these approaches, quite a part of them are based on the idea of ``ghost cell'', such as the methods of Gibou et al.~\cite{gibou2002second}, Majumdar et al.~\cite{majumdar2001rans}, Tseng and Ferziger \cite{tseng2003ghost}, Ghias et al.~\cite{ghias2007sharp}, Mittal et al.~\cite{mittal2008versatile} and others. One problem for the ghost-cell approach is that when dealing with a convex boundary, a ghost cell may be adjacent to more than one fluid cells and there may be several boundary segments lie in more than one directions of the ghost cell. In this singular case, most of the past methods will construct the ghost cell according to the nearest boundary segment. However, it is not rational to assume a unique construction for the ghost cell in such a case. In fact, in our work, some suspicious evidence for this singularity problem can be observed (detailed in Subsection \ref{subsec:tayc}).

In this paper, we propose an IB method based on the ghost-cell approach to simulate the incompressible/compressible viscous flow with stationary/moving boundary. The ghost cell is identified and constructed by an idea of ``local boundary determination'', which allows more than one construction for a sole ghost cell, thus no singularity occurs for the ghost cell. The conception of the ``image point'' presented in some previous methods \cite{majumdar2001rans,tseng2003ghost,ghias2007sharp} is adopted to calculate the variables of the ghost cell. When the boundary is moving, a simple temporal extrapolation is used to settle the so-called ``fresh-cell'' problem \cite{mittal2008versatile}. Test cases show that the method is at least second-order accurate in space globally and locally.

Moreover, we employ the gas-kinetic BGK scheme proposed by Xu \cite{xu2001gas} to solve the flow field. This numerical scheme is based on the gas-kinetic theory and constructed in a more physical way than the Navier-Stokes-function-based scheme. In contrast to the continuum assumption, the physical model of the gas-kinetic scheme is more basic and can describe highly nonequilibrium flow, leading to several advantages of this scheme: robust, positivity-preserving and satisfying the entropy condition spontaneously. In smooth flow region the scheme can attain high accuracy while in discontinuous region it can automatically introduce proper viscosity by a physical mechanism so that the discontinuity can be resolved in the resolution of the mesh. Overall, we adopt this scheme because that it is applicable to the viscous flow from incompressible to compressible, especially suitable for the flow involving both strong discontinuity and shear layer, without any complex artificial fixing.

\section{Numerical method}\label{sec:method}
In this section, we first sketch the gas-kinetic BGK scheme used to govern the flow in our numerical simulation. Then, we describe the grid adopted in our work. Finally, the concept and the construction of our IB method are detailed. It is noteworthy that although the current numerical method is presented in 2D framework, it is not laborious to extend it to 3D situation.

\subsection{Gas-kinetic BGK scheme}\label{subsec:bgk}
The classical gas-kinetic BGK scheme \cite{xu2001gas} is based on finite-volume framework, in which the whole flow field will be divided into control volumes. As with the traditional finite-volume scheme, the governing equations for the BGK scheme can be written as
\begin{equation}\label{eq:gov}
\frac{\partial }{{\partial t}}\int_\Omega  {\vec W } d\Omega  + \oint_{\vec S } {\vec F  \cdot d\vec S }  = 0,
\end{equation}
where $\Omega$ is the control volume and $\vec S$ is the volume surface. $\vec W$ is the vector of conserved variables, defined as
\begin{equation}
\vec W \left( {\vec x ,t} \right) = \mathop {\left( {\rho ,\rho U ,\rho V ,\rho E} \right)}\nolimits^T,
\end{equation}
where $\rho $, $\rho U $, $\rho V $ and $\rho E$ are the mass, momentum and energy densities. $\vec F $ is the flux vector. In 2D cartesian grid, Eq.~\ref{eq:gov} can be discretized as
\begin{equation}\label{eq:discr_gov}
\vec W_{i,j}^{n + 1} = \vec W_{i,j}^n - \frac{1}{A_{i,j}}\int_0^{\Delta t} {\sum\limits_{k = 1}^m {\left( {{{\vec F}_k}\cdot{{\vec S}_k}} \right)} }  ,
\end{equation}
where $\vec W_{i,j}^n$ denotes the average quantity on the cell $(i,j)$ and is assumed to be located at the cell center to the second-order spatial accuracy. $A_{i,j}$ is the cell area, ${{{\vec F }_k}}$ is the flux at the $k$th interface of the cell, ${{{\vec S }_k}}$ is the length of the interface and $m$ is the number of the cell's interfaces.

The issue of a finite-volume scheme is to calculate the flux ${{{\vec F }_k}}$ at the cell interface. To achieve this point, the scheme with accuracy higher than first order will begin with a reconstruction procedure in which the flow variables around the interface will be constructed. In the classical BGK scheme \cite{xu2001gas}, linear reconstruction is employed and the conserved variable ${\vec W'_{i,j}}\left( {x,y} \right)$ inside the cell $(i,j)$ is constructed as
\begin{equation}
{\vec W'_{i,j}}\left( {x,y} \right) = {\vec W_{i,j}} + L_{i,j}^x\left( {x - {x_{i,j}}} \right) + L_{i,j}^y\left( {y - {y_{i,j}}} \right),
\end{equation}
where $L_{i,j}^x$ and $L_{i,j}^y$ are the slopes of the variable in $x$ and $y$ directions respectively. Here, in order to adapt the scheme to the flow with shock, the van Leer limiter is used, i.e.
\begin{equation}
L_{i,j}^x = {\rm{S}}\left( {s_{i,j}^ + ,s_{i,j}^ - } \right)\frac{{\left| {s_{i,j}^ + } \right|\left| {s_{i,j}^ - } \right|}}{{\left| {s_{i,j}^ + } \right| + \left| {s_{i,j}^ - } \right|}},
\end{equation}
where
\begin{eqnarray}
&{s_{i,j}^ +} = {{\left( {{{\vec W}_{i + 1,j}} - {{\vec W}_{i,j}}} \right)} \mathord{\left/
 {\vphantom {{\left( {{{\vec W}_{i + 1,j}} - {{\vec W}_{i,j}}} \right)} {\left( {{x_{i + 1,j}} - {x_{i,j}}} \right)}}} \right.
 \kern-\nulldelimiterspace} {\left( {{x_{i + 1,j}} - {x_{i,j}}} \right)}},\label{eq:slope2}\\
&{s_{i,j}^ -} = {{\left( {{{\vec W}_{i,j}} - {{\vec W}_{i - 1,j}}} \right)} \mathord{\left/
 {\vphantom {{\left( {{{\vec W}_{i,j}} - {{\vec W}_{i - 1,j}}} \right)} {\left( {{x_{i,j}} - {x_{i - 1,j}}} \right)}}} \right.
 \kern-\nulldelimiterspace} {\left( {{x_{i,j}} - {x_{i - 1,j}}} \right)}},\label{eq:slope}\\
&{\rm{S}}\left( {{s_{i,j}^ + },{s_{i,j}^ - }} \right) = {\rm{sign}}\left( {{s_{i,j}^ + }} \right){\rm{ + sign}}\left( {{s_{i,j}^ - }} \right).
\end{eqnarray}
When implementing this reconstruction in a nonuniform Cartesian grid (detailed later in Subsection~\ref{subsec:grid}), the procedure is slightly different. Take the grid in Fig.~\ref{fig:grid_b} as an example, for the cell $(i,j)$, because the cell $(i-1,j)$ is   divided into four cells, when calculating $s_{i,j}^ -$ by Eq.~\ref{eq:slope}, ${\vec W}_{i - 1,j}$ will be replaced by the average value of the four cells, i.e.
\begin{equation}
{\vec W_{i - 1,j}} = \frac{1}{4}\left( {{{\vec W}_{i - 5/4,j + 1/4}} + {{\vec W}_{i - 5/4,j - 1/4}} + {{\vec W}_{i - 3/4,j + 1/4}} + {{\vec W}_{i - 3/4,j - 1/4}}} \right).
\end{equation}
By this means, $s_{i,j}^ -$ will be a second-order estimation of the variable slope at center of the left interface of the cell $(i,j)$. Besides, for the cell $(i - 3/4,j + 1/4)$, when calculating the slope $s_{i - 3/4,j + 1/4}^ +$ at its right interface, due to the absence of the proper adjacent cell, Eq.~\ref{eq:slope2} can not be used. Here, it is calculated by modifying the value of $s_{i,j}^ -$,
\begin{equation}
s_{i - 3/4,j + 1/4}^ +  = s_{i,j}^ -  + \frac{1}{2}\frac{{\partial s}}{{\partial y}} \cdot \Delta {x_{i - 3/4,j + 1/4}},
\end{equation}
where $\Delta {x_{i - 3/4,j + 1/4}}$ is the size of the cell $(i - 3/4,j + 1/4)$ and $\partial s/\partial y$ can be obtained from slopes at left interfaces of the cell $(i - 3/4,j + 1/4)$ and $(i - 3/4,j - 1/4)$, i.e.
\begin{equation}
\frac{{\partial s}}{{\partial y}} = \frac{{s_{i - 3/4,j + 1/4}^ -  - s_{i - 3/4,j - 1/4}^ - }}{{\Delta {x_{i - 3/4,j + 1/4}}}}.
\end{equation}
Thus, all of the slopes can be determined in a second-order manner. In the $y$ direction, the slope $L_{i,j}^y$ can be obtained by a same strategy.

After we have constructed the flow around the interface, the flux going across the interface is to be calculated. As a gas-kinetic scheme, distinguishing from the traditional NS solver, the BGK scheme takes gas particles as carriers of the conserved quantities. A distribution function $f$ is used to describe the particles' dynamics and the flux is obtained by integrating the particles going across the interface. Let the subscript $i+1/2,j$ denote the cell interface of the cell $(i,j)$ in the positive $x$ direction, the relation between the interface flux vector $\vec F_{i + 1/2,j}$ and the distribution function $f$ is
\begin{equation}\label{eq:f2flux}
{\vec F_{i + 1/2,j}} = \int {uf({{\vec x}_{i + 1/2,j}},u,v,\xi ,t)\vec \psi d\Xi } ,
\end{equation}
where $u$ and $v$ are the particle velocities in $x$ and $y$ directions respectively. $\xi $ is the internal variable which can be regarded as a vector with $K$ degrees of freedom. For 2D flow, the particle motion in $z$ direction is included into $\xi $ and $K$ is equal to ${{\left( {5 - 3\gamma } \right)} \mathord{\left/
 {\vphantom {{\left( {5 - 3\gamma } \right)} {\left( {\gamma  - 1} \right)}}} \right.
 \kern-\nulldelimiterspace} {\left( {\gamma  - 1} \right)}} + 1$ \cite{xu1998gas}. $\vec \psi$ is the vector of moments
 \begin{equation}\label{eq:momt}
\vec \psi  = {\left( {{\psi _1},{\psi _2},{\psi _3},{\psi _4}} \right)^T} = {\left( {1,u,v,\frac{1}{2}\left( {{u^2} + {v^2} + {\xi ^2}} \right)} \right)^T},
\end{equation}
$d\Xi  = du dv d\xi $ is the volume element in the velocity space. For more descriptions about the concepts of these variables please refer to Xu \cite{xu2001gas}.

\begin{figure}
\centering
\includegraphics[width=0.45\textwidth]{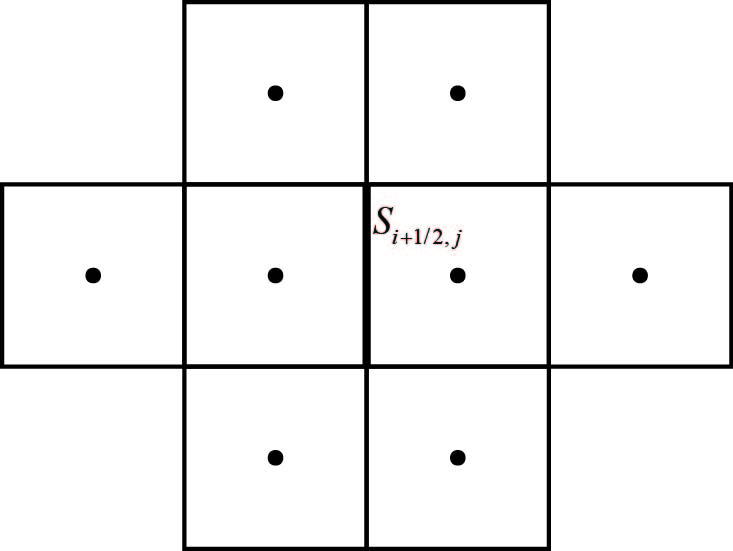}
\caption{\label{fig:stencil} Stencil of the classical BGK scheme when calculating the flux at interface $S_{i+1/2,j}$.}
\end{figure}

So the issue is to calculate the gas distribution function $f$ at the cell interface. In the BGK scheme, the evolution of the distribution function $f$ is governed by the BGK equation \cite{bhatnagar1954model}, whose general solution can be easily obtained. In the classical construction of this scheme, utilizing the Taylor expansion and the Chapman-Enskog expansion, the general solution of the BGK equation is expanded at the interface as \cite{xu2001gas,xu2005multidimensional}
\begin{equation}\label{eq:f_expansion}
\begin{aligned}
f\left( {{{\vec x}_{i + 1/2,j}},u,v,\xi ,t} \right) =  & \left( {1 - {e^{ - t/\tau }}} \right)\left( {1 - t\left( {u{{\bar a}^{lr}} + v\bar b} \right)} \right){g_0}\\&
 - \left( {1 - {e^{ - t/\tau }}} \right)\tau \left( {u{{\bar a}^{lr}} + v\bar b + \bar A} \right){g_0}\\&
 + t\left( {u{{\bar a}^{lr}} + v\bar b + \bar A} \right){g_0}\\&
 + {e^{ - t/\tau }}\left( {1 - t\left( {u{a^{lr}} + v{b^{lr}}} \right)} \right){g^{lr}}\\&
 - {e^{ - t/\tau }}\tau \left( {u{a^{lr}} + v{b^{lr}} + {A^{lr}}} \right){g^{lr}}.
\end{aligned}
\end{equation}
Here, different from the previous papers \cite{xu2001gas,xu2005multidimensional}, we have rearranged the expression of $f$ to make the physical meaning clearer. In Eq.~\ref{eq:f_expansion}, the superscript $lr$ denotes that the variable is calculated in an upwind manner and will be obtained from variables on different sides of the interface depending on the particle velocity $u$. ${g_0},{g^{lr}}$ are Maxwellian distributions and can be got from the conservative variables at both sides of the interface. Other variables, such as ${\bar a^{lr}},\bar b,\bar A$, etc, corresponding to the spacial or temporal derivatives, can be obtained from the slopes of the conserved variables at both sides of the interface. More details about these variables can be found in Xu \cite{xu2001gas} and Xu et al.~\cite{xu2005multidimensional}. Anyhow, surrounding the interface $S_{i+1/2,j}$, $8$ cells' information is required in the above calculation and the stencil is shown in Fig.~\ref{fig:stencil}.

Since we have obtained the gas distribution function $f$ at the interface, the numerical flux across the interface can be computed by Eq.~\ref{eq:f2flux}, and then the conserved variables can be updated through Eq.~\ref{eq:discr_gov}.

At the end of this subsection, we'd like to give some discussion on the above scheme. In Eq.~\ref{eq:f_expansion}, terms multiplied by $\left( {1 - {e^{ - t/\tau }}} \right)$ represent the initial gas distribution while terms multiplied by $e^{ - t/\tau }$ represent the near-equilibrium gas distribution. Terms multiplied by $\tau $ are the nonequilibrium parts of the initial and near-equilibrium gas distribution, which are involved with the viscosity and heat conduction. Term multiplied by $t$ is the substantial derivative of the equilibrium gas distribution. In the continuum limit, $\Delta t \gg \tau $, the gas distribution soon approaches the near-equilibrium state and Eq.~\ref{eq:f_expansion} can be reduced to
\begin{equation}\label{eq:ns_f}
f\left( {{{\vec x}_{i + 1/2,j}},u,v,\xi ,t} \right) = {g_0} - \tau \left( {u{{\bar a}^{lr}} + v\bar b + \bar A} \right){g_0} + t\bar A{g_0},
\end{equation}
which will essentially yield a NS flux \cite{xu2001gas}. It is good enough to use Eq.~\ref{eq:ns_f} to calculate the gas distribution at the interface if the whole flow field is sufficiently smooth. On the other hand, for the flow with discontinuity, the classical BGK scheme introduces an artificial viscosity through
\begin{equation}
\tau  = {\tau _{phys}} + \frac{{\left| {{p_l} - {p_r}} \right|}}{{\left| {{p_l} + {p_r}} \right|}}\Delta t,
\end{equation}
where $p_l,p_r$ are pressures on two sides of the interface and $\Delta t$ is the CFL time step. ${\tau _{phys}}$ is the physical collision time corresponding to the real physical viscosity while the rest part on the right of the equal sign corresponds to the artificial viscosity. In the flow region which is continuum and smooth, the above artificial viscosity will be negligible and a NS flux (Eq.~\ref{eq:ns_f}) will be obtained. In the discontinuous region, the collision time ${\tau}$ will be increased artificially to the scale of $\Delta t$ and a highly nonequilibrium gas distribution will be obtained through Eq.~\ref{eq:f_expansion}, which means that the thickness of the discontinuous layer will be increased to the scale of the mesh size, leading to the non-oscillating resolving of the discontinuity. This is a well-designed mechanism based on the basic model of the gas-kinetic scheme. The only drawback is that in the case of low cell Reynolds number, the CFL time step $\Delta t$ is limited, causing the deviation of the gas distribution function from Eq.~\ref{eq:ns_f}, which will make the scheme break down.

Another thing we want to mention is the Prandtl number fixing. In the BGK model \cite{bhatnagar1954model}, particles with different velocity share a same relaxation rate, resulting in a constant Pr number. Xu \cite{xu2001gas} directly modified the heat flux to fix the Pr number and a similar idea is adopted here. As stated above, in Eq.~\ref{eq:f_expansion}, terms multiplied by $\tau $ are involved with the heat conduction, hence the heat flux of Eq.~\ref{eq:f_expansion} can be obtained through a similar way described in Xu~\cite{xu2001gas}
\begin{equation}
\begin{aligned}
{q_s} =  &  - \tau \left[ {\left( {1 - {e^{ - t/\tau }}} \right)\int {u\left( {u{{\bar a}^{lr}} + v\bar b + \bar A} \right)\left( {{\psi _4} - {\psi _2}{U_0} - {\psi _3}{V_0}} \right)g_0 d\Xi } } \right.\\&
\left. { + {e^{ - t/\tau }}\int {u\left( {u{a^{lr}} + v{b^{lr}} + {A^{lr}}} \right)\left( {{\psi _4} - {\psi _2}{U^{lr}} - {\psi _3}{V^{lr}}} \right)g^{lr} d\Xi } } \right],
\end{aligned}
\end{equation}
where $U_0,V_0,U^{lr},V^{lr}$ are macroscopic gas velocities obtained from $g_0$ and $g^{lr}$ respectively. Then the energy flux ${F_{{\rm{energy}}}}$ in the vector $\vec F$ can be fixed as
\begin{equation}
{F'_{{\rm{energy}}}} = {F_{{\rm{energy}}}} + (\frac{1}{{\Pr }} - 1){q_s}.
\end{equation}
Here, different from previous works \cite{xu2001gas}, the heat flux $q_s$ is a weighted value calculated from the initial and near-equilibrium gas distributions, which may be more compatible with the general solution Eq.~\ref{eq:f_expansion}.

\subsection{Grid strategy}\label{subsec:grid}
\begin{figure}
\centering
\subfigure[\label{fig:grid_a}]{
\includegraphics[width=0.45\textwidth]{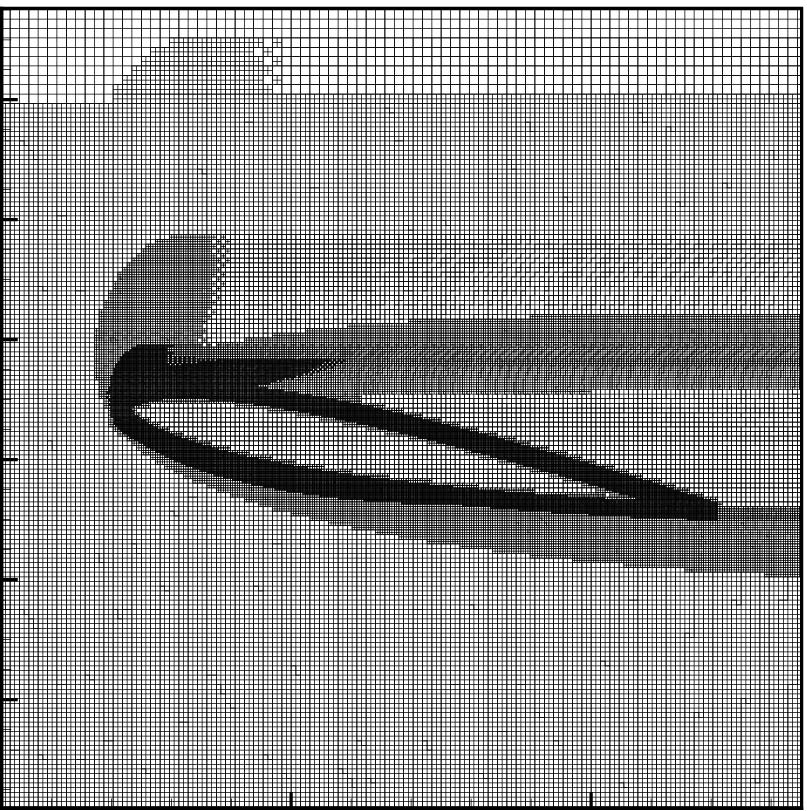}
}\hspace{0.05\textwidth}%
\subfigure[\label{fig:grid_b}]{
\includegraphics[width=0.45\textwidth]{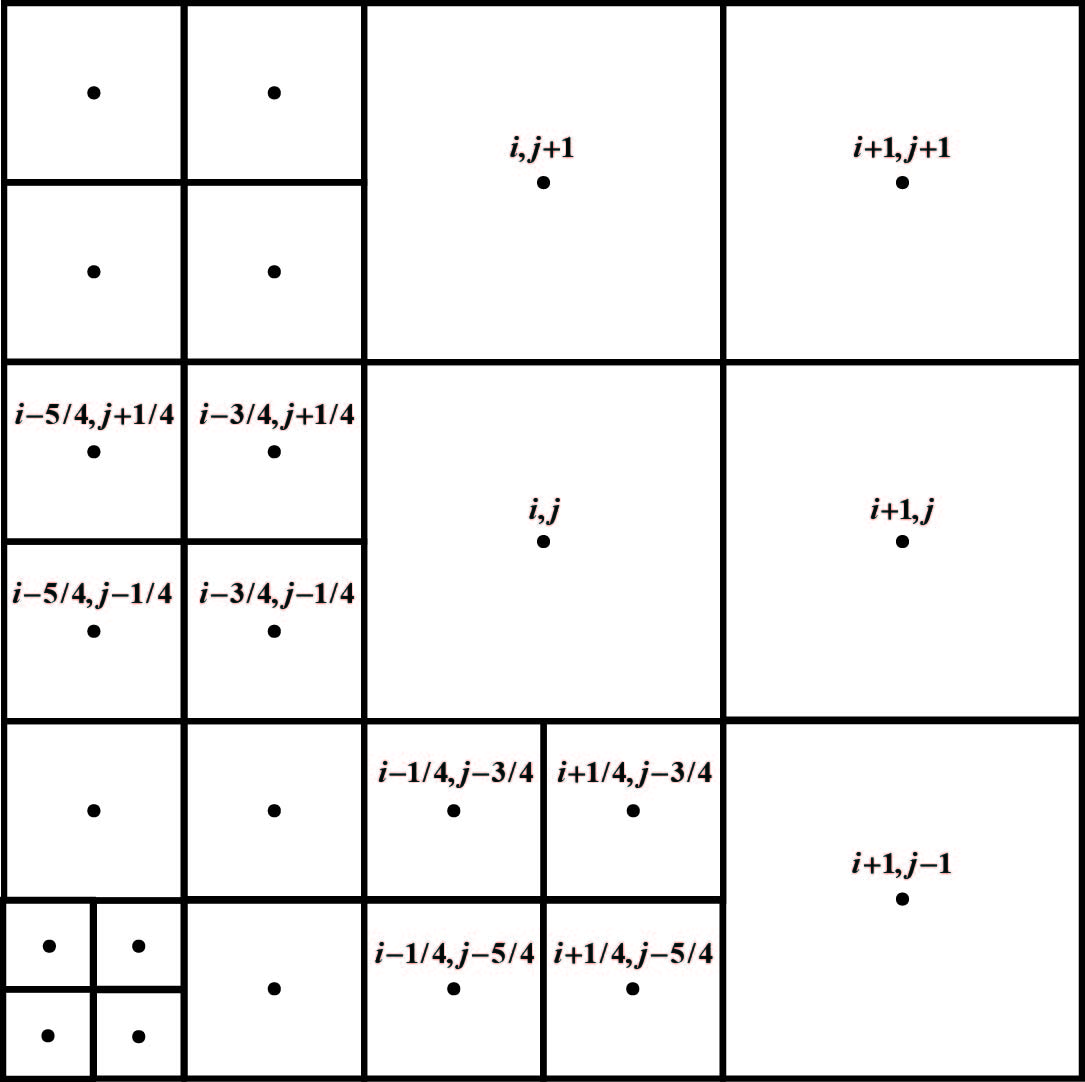}
}
\caption{Cartesian grid with quadtree refinement, (a) the general mesh view for a NACA0012 airfoil with a $10^\circ $ angle of attack and (b) the detailed mesh view. }
\end{figure}

In the IB method, the mesh is not required to fit the solid boundary, thus a non-uniform Cartesian grid is adopted in our work, as shown in Fig.~\ref{fig:grid_a}. To simplify the implementation of the IB method, the grid near the boundary is uniform. A quadtree strategy is used to refine the grid, which means that a cell can be divided into four higher-level subcells while each of the subcells can be subdivided into four cells with a cell level even higher (see Fig.~\ref{fig:grid_b}). In order to avoid too abrupt grid variation and simplify the data structure, cell-level difference greater than one between neighboring cells is not allowed.

The refinement control can be either artificial or solution-adaptive. In the artificial manner, grid refinement and coarsening are managed by a preset procedure. This is applicable to the problem where the boundary motion is predictable. In this case, grid near the boundary will be refined and the refined region is programmed to move with the boundary.
In the solution-adaptive manner, grid refinement and coarsening are controlled by two parameters \cite{de1993quadtree}, i.e.
\begin{equation}
{\tau _c} = \left| {\nabla  \times \vec U} \right|{l^{\frac{{r + 1}}{r}}},
\end{equation}
\begin{equation}
{\tau _d} = \left| {\nabla \cdot\vec U} \right|{l^{\frac{{r + 1}}{r}}},
\end{equation}
where ${\vec U}$ is the fluid velocity, $l$ is the local cell size, $r$ determines the weight of the length scale and is taken to be $2$ in the present work. ${\tau _c}$ and ${\tau _d}$ measure the local strength of the curl and divergence of velocity respectively, which can be used to capture shear layers and shocks \cite{paillere1992wave}. Thus, the refinement criteria is
\begin{equation}
{\tau _c} > {\tau _{c0}}\quad {\rm{or}}\quad {\tau _d} > {\tau _{d0}},
\end{equation}
while the coarsening criteria is
\begin{equation}
{\tau _c} < c{\tau _{c0}}\quad {\rm{and}}\quad {\tau _d} < c{\tau _{d0}},
\end{equation}
where ${\tau _{c0}}$, ${\tau _{d0}}$ and $c$ are user-defined parameters and the value of $c$ should be less than ${\left( {\frac{1}{2}} \right)^{\frac{{r + 1}}{r}}}$ to avoid recursive refinement and coarsening.

In a problem where the boundary is stationary and the flow is steady, the above refinement will be done only a few times while in an unsteady problem such procedures will be executed at intervals.

\subsection{Immersed-boundary method}\label{subsec:ib}
\begin{figure}
\centering
\subfigure[]{
\includegraphics[width=0.45\textwidth]{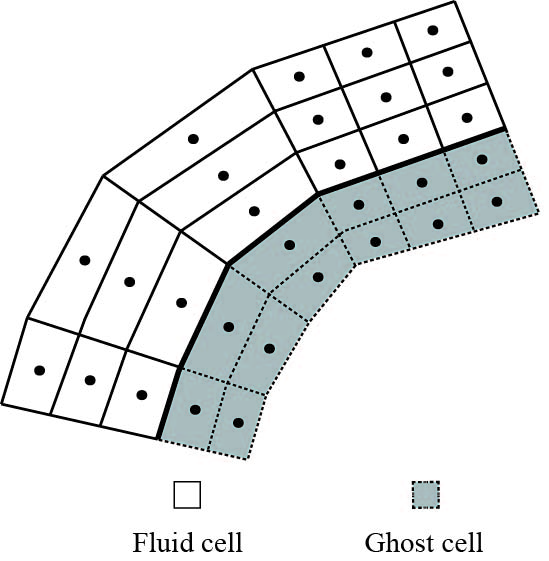}
}\hspace{0.05\textwidth}%
\subfigure[]{
\includegraphics[width=0.45\textwidth]{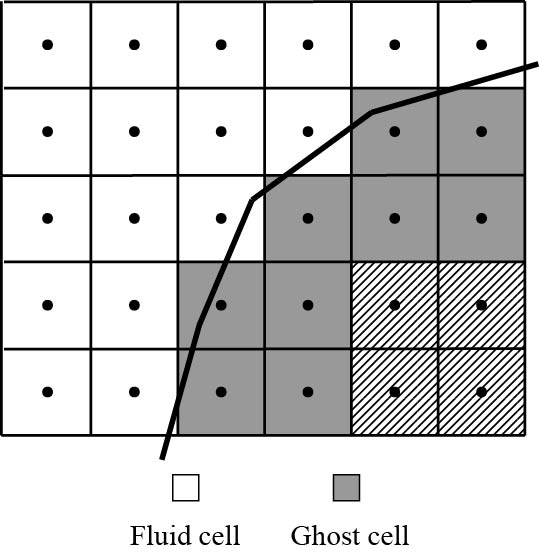}
}
\caption{\label{fig:ghost}Fluid cells and ghost cells in (a) boundary-conforming method and (b) IB method. The heavy polyline is the solid boundary.}
\end{figure}

Forming a mirrored flow field in the solid region to fulfil the boundary condition is applied in plenty of boundary-conforming as well as IB methods \cite{majumdar2001rans,gibou2002second,tseng2003ghost,ghias2007sharp,mittal2008versatile}. As shown in Fig.~\ref{fig:ghost}, in these methods, cells in the flow field are called as ``fluid cells''. To complement the calculation stencils of the fluid cells near the boundary, a layer of cells near the boundary and inside the solid body are constructed as ``ghost cells'' through various of strategies, so that all fluid cells can be updated by a unified scheme.

\begin{figure}
\centering
\includegraphics[width=0.45\textwidth]{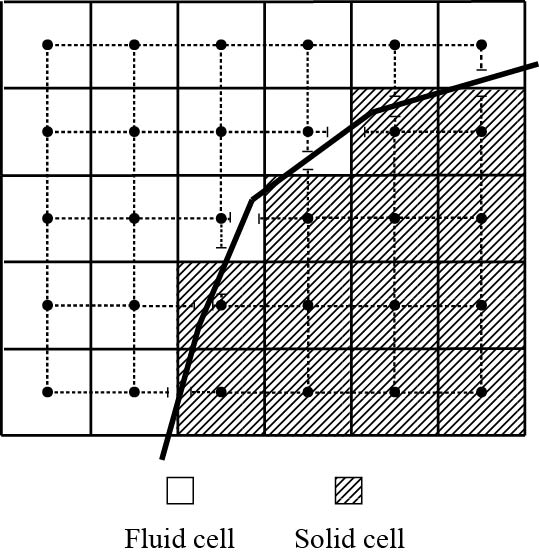}
\caption{\label{fig:linkcell} Determination of the fluid cells and the solid cells. The heavy polyline denotes the boundary segments while the dashed lines denote connections between cell-center nodes.}
\end{figure}

The same idea is applied in our work. In the present method, the boundary is represented by line segments and a uniform Cartesian grid is used around the boundary. The first step is to identify the fluid cells, whose centers are outside the solid body. The rest of the cells, whose centers lie inside the solid body, are defined as ``solid cells''. Although there are various of ways can be used to do this identification including the ray-crossing technique \cite{O1998Computational}, here we adopt a way more compatible with our succedent procedure. As shown in Fig.~\ref{fig:linkcell}, if two cells share a same interface, their cell-center nodes are connected by dashed line. As a result, a topology of the cell-center nodes is formed. Then let the solid boundary cut off the connecting lines between different cell centers. After that we can pick an arbitrary fluid cell, all cells whose center nodes are connected, directly or indirectly, with the center of the picked cell are fluid cells. The boundary segments cutting off the cell-center connecting lines will be recorded and used in the succedent procedure. On the fluid cells, fluid variables will be updated normally by the numerical scheme, while the solid cells will be let alone and not involved in the calculation.

\begin{figure}
\centering
\includegraphics[width=0.45\textwidth]{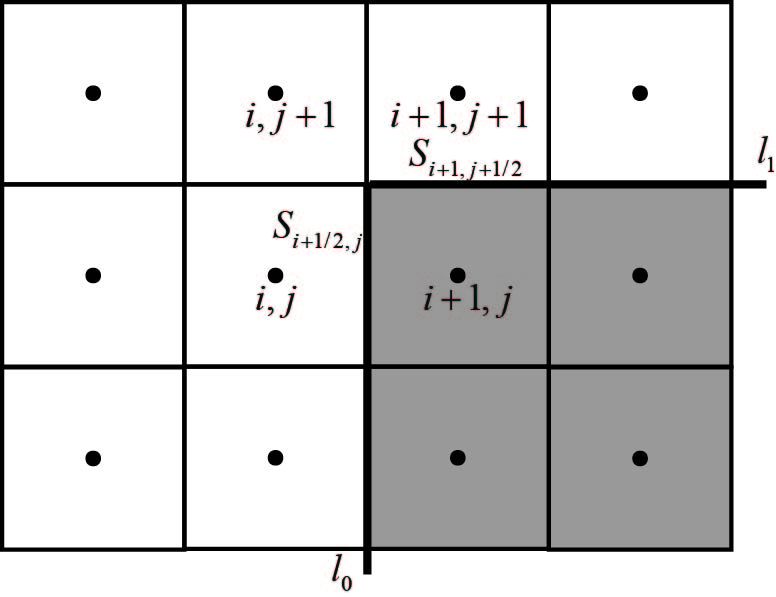}
\caption{\label{fig:case_sp} Singularity problem when determining the ghost cell $(i+1,j)$. $l_0,l_1$ are two segments of the solid boundary. $S_{i+1/2,j}$ and $S_{i+1,j+1/2}$ are  interfaces of the ghost cell $(i+1,j)$. Singularity occurs when constructing the ghost cell $(i+1,j)$ by the boundary segment $l_0$ or $l_1$.}
\end{figure}

In the next step, the ghost cells will be identified and constructed by a novel strategy. In the previous ghost-cell IB methods \cite{majumdar2001rans,tseng2003ghost,ghias2007sharp,mittal2008versatile}, the ghost cell will be uniquely determined and variables on a ghost cell will have unique values for all interfaces of the cell, which may suffer singularity problem in some cases. As shown in Fig.~\ref{fig:case_sp}, two boundary segments, $l_0$ and $l_1$, with different angles, lie in two directions of the ghost cell $(i+1,j)$ which can be constructed based on either of the two segments. It is not rational to assume a unique construction for ghost cell $(i+1,j)$ in such a case. A reasonable practice is using the information of $l_0$ to construct $(i+1,j)$ when calculating the variables at interface $S_{i+1/2,j}$, while using the information of $l_1$ to do the construction when dealing with the interface $S_{i+1,j+1/2}$, just as what is done in a conventional body-fitted grid. Further more, in the conventional body-fitted ghost-cell treatment, when calculating the flux at the interface $S_{i+1/2,j}$, the fluid cell $(i+1,j+1)$ which is in the calculation stencil will be regarded as a ghost cell on which the variables will be constructed by the information of $l_0$ and $(i,j+1)$, as if the boundary segment $l_0$ were extended between the cells $(i,j+1)$ and $(i+1,j+1)$. Indeed, the flow passes through the interface $S_{i+1/2,j}$ will meet the boundary segment $l_0$, which will exert a reaction on the flow to fulfill the boundary condition. Hence the local boundary segment $l_0$ should determine the construction of the ghost cells when calculating the variables at $S_{i+1/2,j}$.

\begin{figure}
\centering
\includegraphics[width=0.45\textwidth]{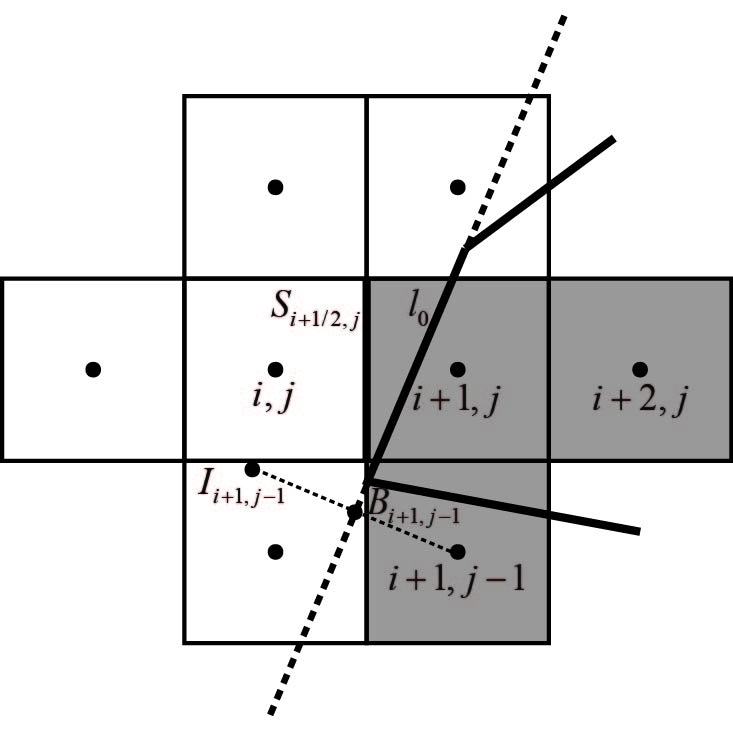}
\caption{\label{fig:ghost_identify} Determination of the ghost cells. The boundary segment $l_0$ will be extended to a virtual boundary line and the gray cells will be regarded as ghost cells when calculating the flux at $S_{i+1/2,j}$. $B_{i+1,j-1}$ and $I_{i+1,j-1}$ are the boundary-intercept point and image point for the ghost cell $(i+1,j-1)$ respectively.}
\end{figure}

Thus, a ``local boundary determination'' thought is implemented in this paper to complete the ghost-cell construction. When calculating the variable at a interface between a fluid cell and a solid cell, the local boundary segment cutting off the line connecting the two cell centers will be extended to a line. Then this line will be regarded as a virtual solid boundary in the vicinity of the interface. All cells in the calculation stencil whose center nodes lie on the solid side of the line will be identified as ghost cells. Just as what is shown in Fig.~\ref{fig:ghost_identify}, when calculating the flux at the interface $S_{i+1/2,j}$, the calculation stencil includes $8$ cells shown in the picture, where $2$ cells, $(i+1,j)$ and $(i+2,j)$, are solid cells while the rest $6$ cells are fluid cells. The boundary segment $l_0$ which cuts off the connecting line between the cells $(i,j)$ and $(i+1,j)$ will be extended to a virtual boundary line. The solid cells $(i+1,j)$, $(i+2,j)$ and the fluid cell $(i+1,j-1)$ whose centers lie on the solid side of the extended $l_0$ will be identified as ghost cells.

After that, the conception of image point \cite{majumdar2001rans,tseng2003ghost,ghias2007sharp,mittal2008versatile} is introduced here to construct the variable at the center of the ghost cell. As shown in Fig.~\ref{fig:ghost_identify}, for the ghost cell $(i+1,j-1)$, the point $I_{i+1,j-1}$ which is symmetric with the cell-center node about the extended boundary segment $l_0$ is the image point. The line connecting these two symmetric points intersects the extended $l_0$ at the boundary-intercept point $B_{i+1,j-1}$, where the boundary conditions for $l_0$ should be satisfied. The boundary condition can be either Dirichlet or Neumann type. For a Dirichlet boundary condition which has the form
\begin{equation}
{\phi _B} = {\phi _0},
\end{equation}
where $\phi _B$ is a generic variable at the boundary-intercept point and $\phi _0$ is the given value for it, the variable $\phi _{i + 1,j-1}$ at the center node of $(i+1,j-1)$ can be calculated as
\begin{equation}
{\phi _{i + 1,j-1}} = 2{\phi _0} - {\phi _I},
\end{equation}
where $\phi _I$ is the corresponding variable at the image point $I_{i+1,j-1}$. For a Neumann boundary condition, i.e.
\begin{equation}
{\vec n_0} \cdot {(\nabla \phi )_B} = {\left( {\frac{{\partial \phi }}{{\partial n}}} \right)_0},
\end{equation}
where $\vec n_0$ is the outer normal vector of $l_0$, $(\nabla \phi )_B$ is the gradient at the boundary-intercept point and ${\left( {\frac{{\partial \phi }}{{\partial n}}} \right)_0}$ is the given normal gradient, $\phi _{i + 1,j-1}$ can be calculated as
\begin{equation}
{\phi _{i + 1,j-1}} = {\phi _I} - {d_I}{\left( {\frac{{\partial \phi }}{{\partial n}}} \right)_0},
\end{equation}
where ${d_I}$ is the distance between $I_{i + 1,j-1}$ and the center of $(i + 1,j-1)$. Therefore, once we have obtained the variable $\phi _I$ at the image point $I_{i+1,j-1}$, the variable for the ghost cell $(i+1,j-1)$ can be determined by a linear construction.

\begin{figure}
\centering
\subfigure[\label{fig:image_a}]{
\includegraphics[width=0.45\textwidth]{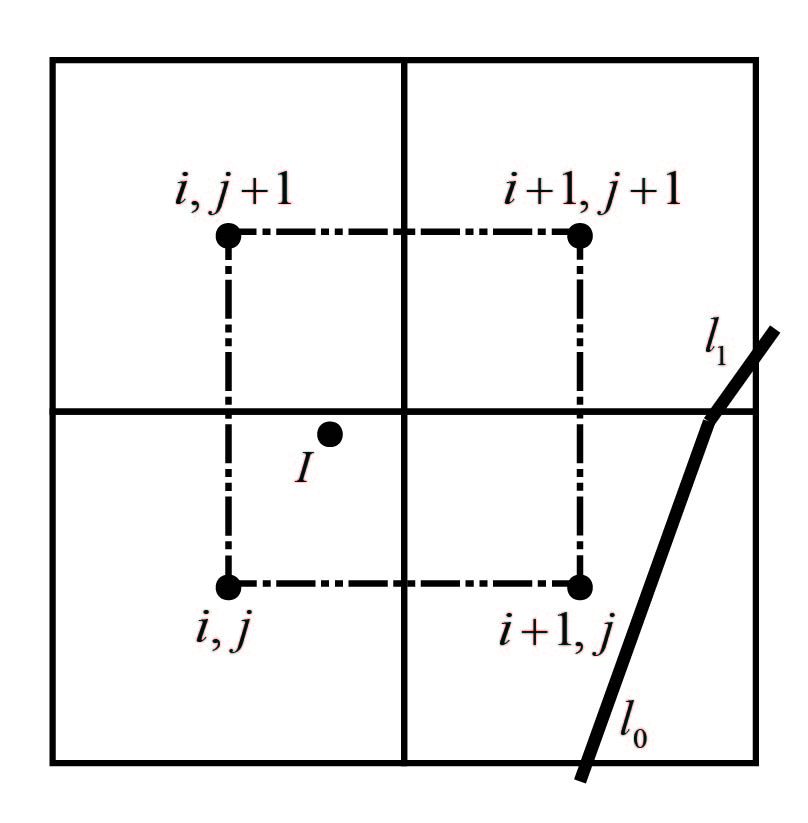}
}\hspace{0.05\textwidth}%
\subfigure[\label{fig:image_b}]{
\includegraphics[width=0.45\textwidth]{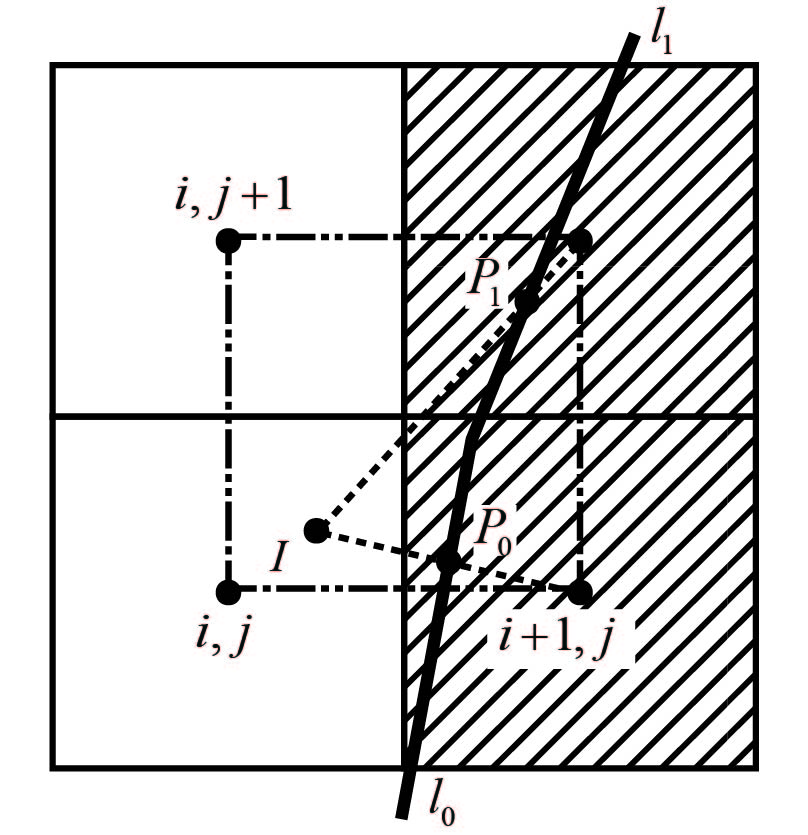}
}
\caption{\label{fig:image}Cases may be encountered when interpolating the variable at the image point $I$ by bilinear interpolation, (a) surrounding $4$ cells are all fluid cells and (b) $2$ of the surrounding cells are solid cells.}
\end{figure}

Like in the predecessor's works \cite{ghias2007sharp,mittal2008versatile}, a bilinear interpolation is adopted to obtain the value of the variable at the image point. The interpolation has the form
\begin{equation}\label{eq:bilinear}
\phi  = {c_0} + {c_1}x + {c_2}y + {c_3}xy,
\end{equation}
where $c_0$$\sim$$c_3$ are $4$ unknown coefficients which can be determined by $4$ interpolation conditions. In this step we try to get the real value of the fluid variable at the image point, so the virtual boundary, which is only used for ghost cell determination, is put aside and the information of the real solid boundary is taken into account. Several cases will be encountered as shown in Fig.~\ref{fig:image}. The simplest case is that $4$ cells encircling the image point are fluid cells, like what is depicted in Fig.~\ref{fig:image_a}. In this case, directly substituting the information of the $4$ fluid cells into Eq.~\ref{eq:bilinear} will yield
\begin{equation}\label{eq:image_fun_0}
\left( {\begin{array}{*{20}{c}}
1&{{x_{i,j}}}&{{y_{i,j}}}&{{x_{i,j}}{y_{i,j}}}\\
1&{{x_{i + 1,j}}}&{{y_{i + 1,j}}}&{{x_{i + 1,j}}{y_{i + 1,j}}}\\
1&{{x_{i,j + 1}}}&{{y_{i,j + 1}}}&{{x_{i,j + 1}}{y_{i,j + 1}}}\\
1&{{x_{i + 1,j + 1}}}&{{y_{i + 1,j + 1}}}&{{x_{i + 1,j + 1}}{y_{i + 1,j + 1}}}
\end{array}} \right)\left( \begin{array}{l}
{c_0}\\
{c_1}\\
{c_2}\\
{c_3}
\end{array} \right) = \left( \begin{array}{l}
{\phi _{i,j}}\\
{\phi _{i + 1,j}}\\
{\phi _{i,j + 1}}\\
{\phi _{i + 1,j + 1}}
\end{array} \right),
\end{equation}
from which the $4$ unknown coefficients can be solved out and the variable at the image point can be determined. The situation will be more complicated when $1$ or $2$ of the surrounding $4$ cells are solid cells. In this case, a line will be drawn between the image point and the center of the solid cell, which will intersect the boundary segment at a point where the boundary condition can be used to close the interpolation. Just as what is shown in Fig.~\ref{fig:image_b}, the line segments between the image point $I$ and the center nodes of the solid cells intersect the boundary segments $l_0$ and $l_1$ at $P_0$ and $P_1$ respectively. If a Dirichlet boundary condition is applied at $P_0$ and $P_1$, a function similar to Eq.~\ref{eq:image_fun_0} can be got where the variables corresponding to the solid cells $(i+1,j)$ and $(i+1,j+1)$ will be replaced by the corresponding values at $P_0$ and $P_1$. If a Neumann boundary condition is applied, i.e.
\begin{equation}\label{eq:neu}
\begin{aligned}
{\vec n_0} \cdot {(\nabla \phi )_{{P_0}}} = {\left( {\frac{{\partial \phi }}{{\partial n}}} \right)_0},\\
{\vec n_1} \cdot {(\nabla \phi )_{{P_1}}} = {\left( {\frac{{\partial \phi }}{{\partial n}}} \right)_1},
\end{aligned}
\end{equation}
where ${\vec n_0} = {({n_{0x}},{n_{0y}})^T}$ and ${\vec n_1} = {({n_{1x}},{n_{1y}})^T}$ are the outer normal vectors of $l_0$ and $l_1$, taking the derivative of  Eq.~\ref{eq:bilinear} and substituting it into the boundary conditions Eq.~\ref{eq:neu}, and then combining the interpolation conditions at the $2$ fluid cells will yield
\begin{equation}
\left( {\begin{array}{*{20}{c}}
1&{{x_{i,j}}}&{{y_{i,j}}}&{{x_{i,j}}{y_{i,j}}}\\
1&{{x_{i,j + 1}}}&{{y_{i,j + 1}}}&{{x_{i,j + 1}}{y_{i,j + 1}}}\\
0&{{n_{0x}}}&{{n_{0y}}}&{{n_{0x}}{y_0} + {n_{0y}}{x_0}}\\
0&{{n_{1x}}}&{{n_{1y}}}&{{n_{1x}}{y_1} + {n_{1y}}{x_1}}
\end{array}} \right)\left( {\begin{array}{*{20}{l}}
{{c_0}}\\
{{c_1}}\\
{{c_2}}\\
{{c_3}}
\end{array}} \right) = \left( {\begin{array}{*{20}{l}}
{{\phi _{i,j}}}\\
{{\phi _{i,j + 1}}}\\
{{{\left( {{{\partial \phi } \mathord{\left/
 {\vphantom {{\partial \phi } {\partial n}}} \right.
 \kern-\nulldelimiterspace} {\partial n}}} \right)}_0}}\\
{{{\left( {{{\partial \phi } \mathord{\left/
 {\vphantom {{\partial \phi } {\partial n}}} \right.
 \kern-\nulldelimiterspace} {\partial n}}} \right)}_1}}
\end{array}} \right),
\end{equation}
where $x_0,y_0$ and $x_1,y_1$ are coordinates of $P_0$ and $P_1$ respectively. Afterwards the $4$ unknown coefficients can be solved out and the value of the variable at the image point can be obtained by the interpolation Eq.~\ref{eq:bilinear}.

Thus, just like what is shown in Fig.~\ref{fig:ghost_identify}, all ghost cells in the calculation stencil can be determined and eventually all fluid cells can be updated by a unified scheme.

\begin{figure}
\centering
\includegraphics[width=0.45\textwidth]{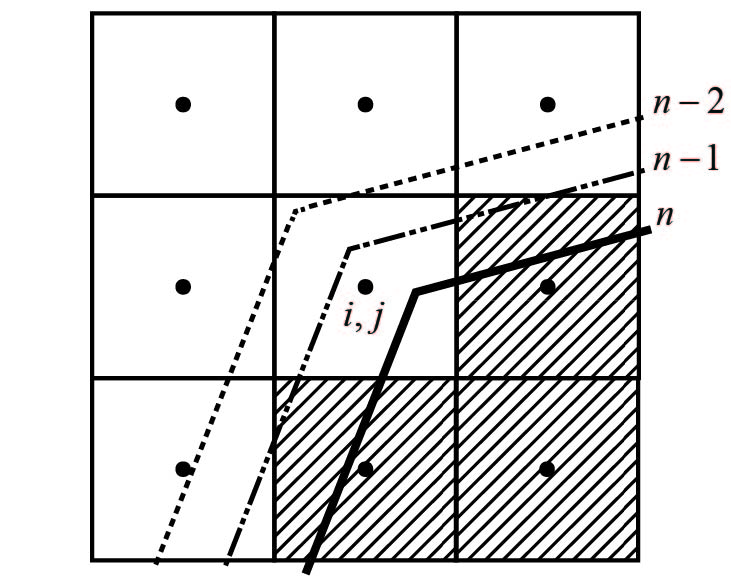}
\caption{\label{fig:freshcell}The fresh-cell problem when implementing the IB method in the simulation with a moving boundary. The dashed line, dot-dash line, and solid line denote the boundaries at the $(n-2)$th, $(n-1)$th and $n$th time steps respectively. The cell $(i,j)$ is a fresh cell at the $n$th step.}
\end{figure}

\begin{figure}
\centering
\subfigure[\label{fig:freshcell_weight_a}]{
\includegraphics[width=0.45\textwidth]{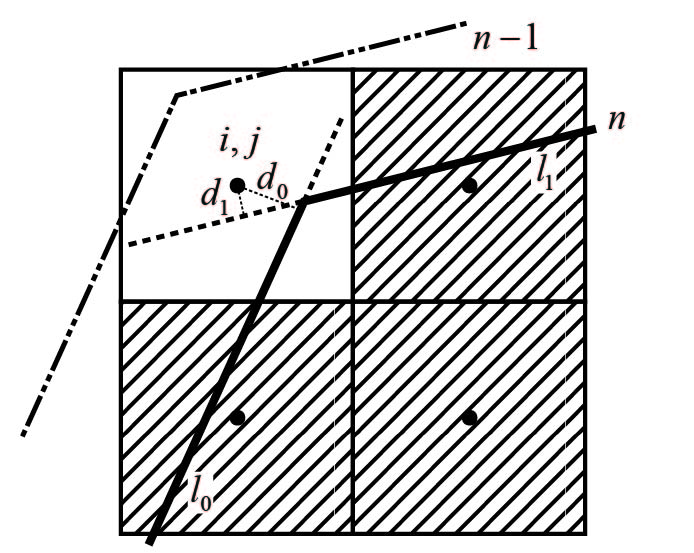}
}\hspace{0.05\textwidth}%
\subfigure[\label{fig:freshcell_weight_b}]{
\includegraphics[width=0.45\textwidth]{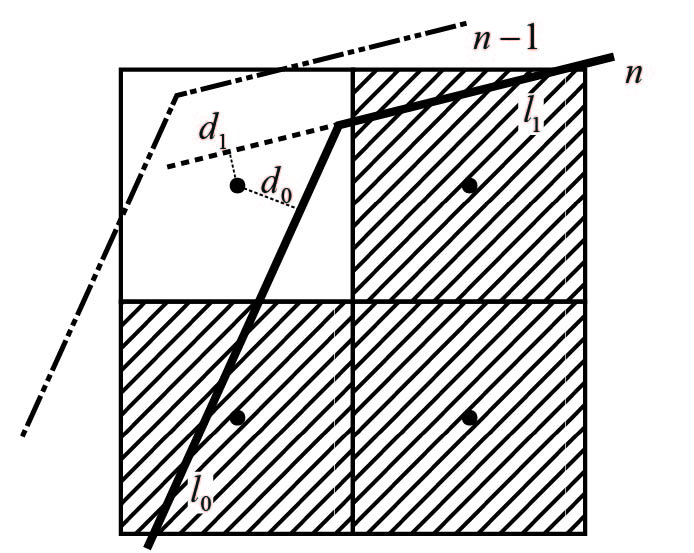}
}
\caption{Weights for different ghost-cell constructions on the fresh cell $(i,j)$. There are $2$ ghost-cell constructions for $(i,j)$ corresponding to the boundary segments $l_0$ and $l_1$ at the $(n-1)$th step. $d_0$, $d_1$ are the normal distances that the center node of $(i,j)$ outside the boundary segments $l_0$, $l_1$ at the $n$th step.}
\end{figure}

When implementing the above method in a simulation where the boundary is moving, the so-called ``fresh-cell'' problem \cite{mittal2008versatile} will be encountered. This happens when a solid cell turning into a fluid cell along with the boundary motion. Just as displayed in Fig.~\ref{fig:freshcell}, the cell $(i,j)$ is a solid cell at the $(n-2)$th, $(n-1)$th steps and turns into a fluid cell at the $n$th step. How to determine the variable on this new fluid cell (i.e. fresh cell) is the issue. Here a simple extrapolation in time is adopted to resolve this problem. A generic variable $\phi _{i,j}^n$ for the fresh cell $(i,j)$ at $n$th step will be calculated as
\begin{equation}\label{eq:fresh_ex}
\phi _{i,j}^n = \frac{{{t^n} - {t^{n - 2}}}}{{{t^{n - 1}} - {t^{n - 2}}}}\phi _{i,j}^{n - 1} + \frac{{{t^{n - 1}} - {t^n}}}{{{t^{n - 1}} - {t^{n - 2}}}}\phi _{i,j}^{n - 2},
\end{equation}
where $t$ is the time and the superscript denotes the number of the time step, $\phi _{i,j}^{n - 1}$ and $\phi _{i,j}^{n - 2}$ are the ghost-cell values of the cell $(i,j)$ at $(n-1)$th, $(n-2)$th steps respectively. It is notable that in the present method, a solid cell may corresponds to more than one ghost cell, whose variables will be constructed from different boundary segments. So $\phi _{i,j}^{n - 1}$ and $\phi _{i,j}^{n - 2}$ in Eq.~\ref{eq:fresh_ex} are obtained in a weighting manner, i.e. if the cell $(i,j)$ at the $(n-1)$th step has $m$ ghost-cell constructions corresponding to the boundary segments $l_0, l_1,...,l_{m-1}$ respectively, then $\phi _{i,j}^{n - 1}$ can be calculated as
\begin{equation}\label{eq:fresh_weight}
\phi _{i,j}^{n - 1} = \frac{{\sum\limits_{k = 0}^{m - 1} {{\rm{H}}[{d_k}]{d_k}\phi _k^{n - 1}} }}{{\sum\limits_{k = 0}^{m - 1} {{\rm{H}}[{d_k}]{d_k}} }},
\end{equation}
where $d_k$ is the normal distance that the center of $(i,j)$ outside the boundary segment $l_k$ at the $n$th step, $\phi _k^{n - 1}$ is the ghost-cell value corresponding to $l_k$, $\rm{H}[x]$ is the Heaviside function defined as
\begin{equation}
{\rm{H}}[x] = \left\{ \begin{array}{l}
0,\quad x < 0;\\
1,\quad x \ge 0.
\end{array} \right.
\end{equation}
The weight for certain ghost-cell value is calculated in consideration of the fact that the center node of the fresh cell will be biased to a more distant (in the sense of normal distance) boundary segment, based on which the ghost-cell value should have a larger weight. As shown in Fig.~\ref{fig:freshcell_weight_a}, the center node of the cell $(i,j)$ is biased to $l_0$ which is more distant than $l_1$ (i.e. ${d_0} > {d_1}$) at the $n$th step, thereby a larger weight will be given to the ghost-cell value constructed from $l_0$. In Fig.~\ref{fig:freshcell_weight_b}, the center node of $(i,j)$ is totally biased to $l_0$ and lies on the solid side of $l_1$ at the $n$th step, which means that the center node never goes over to the fluid side of $l_1$ and is always a point inside the solid body for this boundary segment. In this case a negative $d_1$ will be got and the weight for the ghost-cell value based on $l_1$ is $0$.

Another notable thing is that, just like various of previous IB methods \cite{ghias2007sharp,mittal2008versatile}, the present IB method cannot guarantee the conservative condition near the immersed boundary, which may lead to a persistent increase or decrease in the conserved quantity. Therefore it is necessary to give boundary conditions which can make the problem well-posed. If the flow is well-posed, the conservative problem may become just a matter of accuracy.

\section{Numerical results and discussions}\label{sec:results}
In this section, several test cases are conducted to validate the present method. First, the spacial accuracy of the method is investigated in the simulation of Taylor-Couette flow. Then, the test case of supersonic flow over a stationary circular cylinder is carried out to testify the ability of the present method in handling highly compressible flow. Finally, the flow over an oscillating circular cylinder and the flow over a moving airfoil are conducted to demonstrate the capability of the method in managing the moving boundary.

\begin{figure}
\centering
\includegraphics[width=0.45\textwidth]{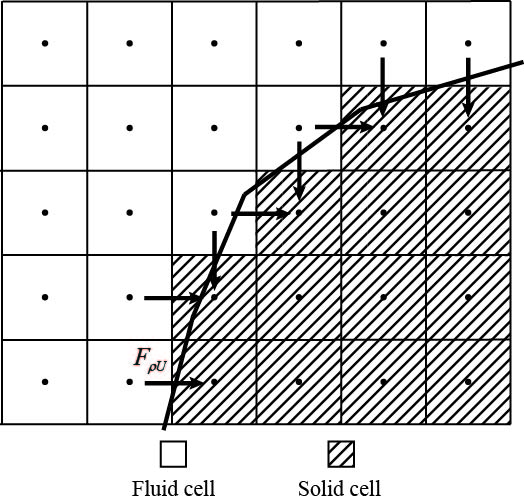}
\caption{\label{fig:cal_force}Calculation of the force exerted on the solid body by fluid. $F_{\rho U}$ is the flux of $x$-momentum.}
\end{figure}

In this part, the force which fluid exerts on the solid body is calculated by summing the momentum fluxes going from the fluid cells into the solid cells. Just as shown in Fig.~\ref{fig:cal_force}, the $x$-direction component of the force $D_x$ can be calculated as
\begin{equation}\label{eq:Dx}
{D_x} = \frac{{\int_0^{\Delta t} {\sum\limits_k {({F_{\rho U,k}} - {U_0}{F_{\rho ,k}})} } {S_k}}}{{\Delta t}},
\end{equation}
where $\Delta t$ is the time step, ${{F_{\rho U,k}}}$ and $F_{\rho ,k}$ are the $x$-momentum and mass fluxes going across the $k$th interface into the solid cell, $U_0$ is the $x$-velocity of the solid body and $S_k$ is the area of the interface.

\subsection{Taylor-Couette flow}\label{subsec:tayc}
\begin{figure}
\centering
\includegraphics[width=0.45\textwidth]{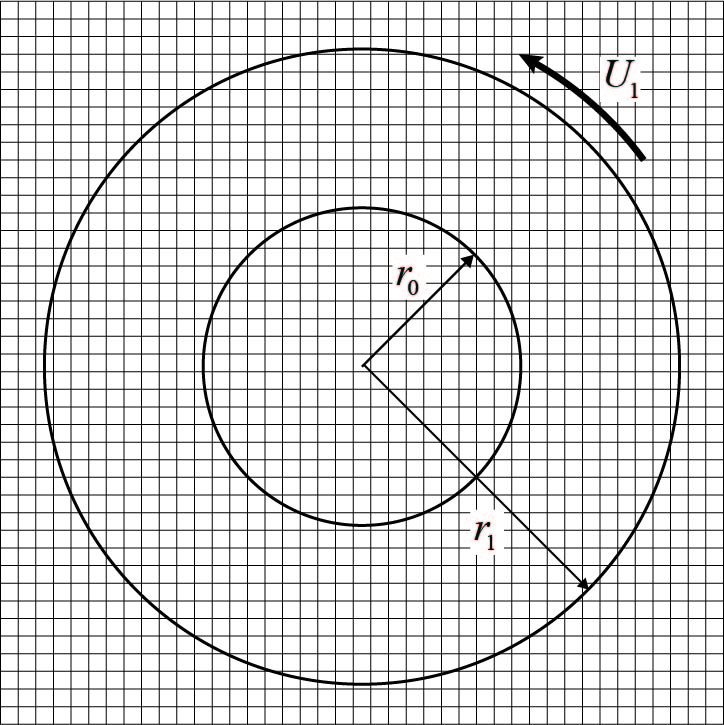}
\caption{\label{fig:test1_taylor}Schematic of the simulation of Taylor-Couette flow.}
\end{figure}

We have simulated the Taylor-Couette flow to assess the spatial accuracy of the current method. In this test case, viscous fluid is filled in the gap of two concentric rotating cylinders, as shown in Fig.~\ref{fig:test1_taylor}. In incompressible limit, an analytical solution can be solved out for this flow \cite{White2005Viscous}, and the variables at radius $r$ can be calculated as
\begin{equation}
{U_\theta } = {C_1}r + \frac{{{C_2}}}{r},
\end{equation}
\begin{equation}
p = \frac{1}{2}\rho \left( {C_1^2{r^2} - \frac{{C_2^2}}{{{r^2}}} + 4{C_1}{C_2}\ln r} \right),
\end{equation}
\begin{equation}
\frac{1}{\lambda } =  - \frac{{2(\gamma  - 1)}}{\gamma }\frac{{\Pr C_2^2}}{{{r^2}}} + {C_3}\ln r + {C_4},
\end{equation}
where $U_\theta$ is the circumferential velocity of the fluid, $p$ is the pressure, $\gamma$ is the ratio of specific heat, $\lambda$ is a variable involved in the gas-kinetic scheme and related to the gas temperature $T$ which can be calculated as $\lambda  = 1/(2RT)$, where $R$ is the specific gas constant. The integration constants $C_1$$\sim$$C_4$ can be determined by the radiuses and the boundary conditions of the inner and outer cylinders, which are omitted here for simplicity.

Our simulation employs a square computational domain with uniform Cartesian grid, as shown in Fig.~\ref{fig:test1_taylor}. The radiuses of the two cylinders are set to be ${r_1} = 2{r_0}$. The inner cylinder is holding stationary while the outer cylinder is rotating at a velocity $U_1$ equal to ${\rm{Ma}} = 0.01$ (approximating the incompressible flow). The Reynolds number defined by the width of the gap $d=r_1-r_0$ and the velocity $U_1$ is ${\rm{Re}} = 25$. The Prandtl number is set to be ${\rm{Pr}} = 1$. To make the problem well-posed, the boundary conditions at the two cylinders are set to be
\begin{equation}
\begin{array}{l}
{\rm{inner}}:\quad {U_0} = 0,\quad \quad {\left( {\frac{{\partial p}}{{\partial n}}} \right)_0} = 0,\quad {\lambda _0} = \frac{\gamma }{2} - 0.0001,\\
{\rm{outer}}:\quad {U_1} = 0.01,\quad {p_1} = \frac{1}{\gamma },\quad \quad \;{\lambda _1} = \frac{\gamma }{2},
\end{array}
\end{equation}
where all of the variables are nondimensionalized by the density and the speed of sound at the outer cylinder. $4$ grid sizes, $\Delta x = d/25,d/35,d/50,d/75$, are adopted to simulate this flow.

\begin{figure}
\centering
\subfigure[]{
\includegraphics[width=0.45\textwidth]{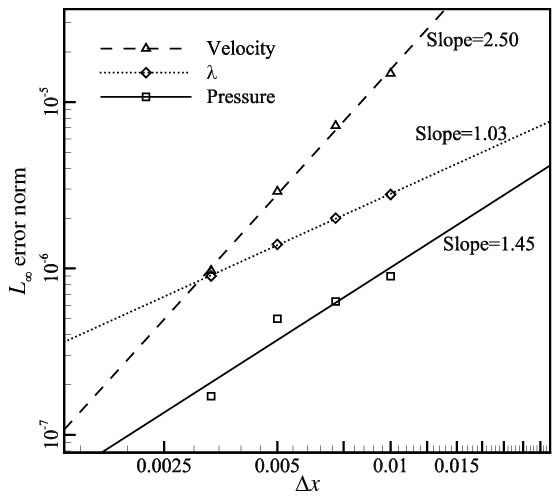}
}\hspace{0.05\textwidth}%
\subfigure[]{
\includegraphics[width=0.45\textwidth]{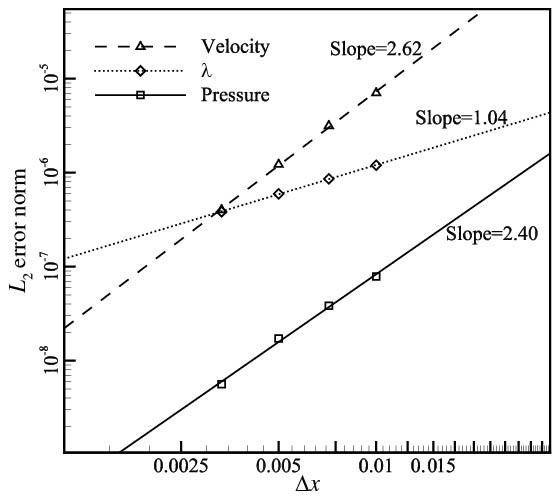}
}
\caption{\label{fig:test1_gksL}(a) ${L_\infty }$ and (b) ${L_2 }$ error norms for the simulation of Taylor-Couette flow using the classical gas-kinetic BGK scheme. Norms are calculated from all fluid cells in the flow field.}
\end{figure}

\begin{figure}
\centering
\subfigure[]{
\includegraphics[width=0.45\textwidth]{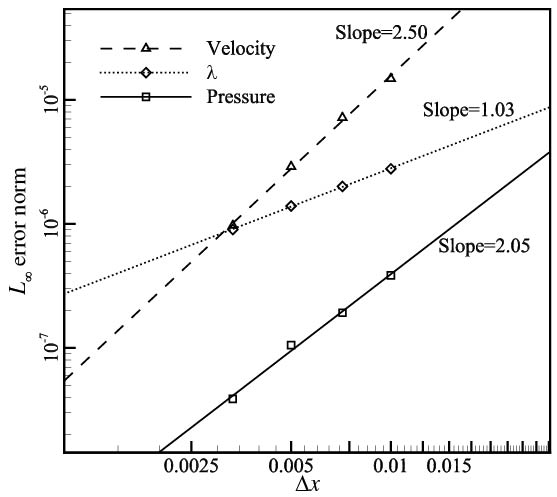}
}\hspace{0.05\textwidth}%
\subfigure[]{
\includegraphics[width=0.45\textwidth]{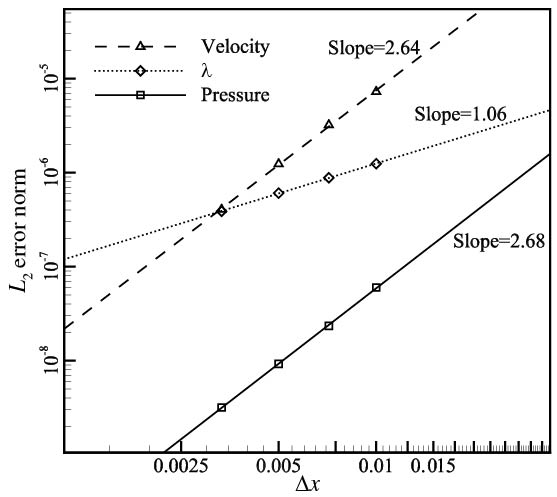}
}
\caption{\label{fig:test1_gksLf}(a) ${L_\infty }$ and (b) ${L_2 }$ error norms for the simulation of Taylor-Couette flow using the classical gas-kinetic BGK scheme. Norms are calculated from all fluid cells except those adjacent to solid cells.}
\end{figure}

\begin{figure}
\centering
\subfigure[\label{fig:test1_gks_pre}]{
\includegraphics[width=0.45\textwidth]{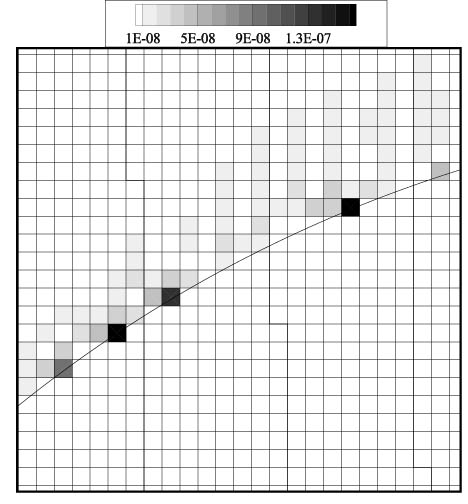}
}\hspace{0.05\textwidth}%
\subfigure[\label{fig:test1_gks_lambda}]{
\includegraphics[width=0.45\textwidth]{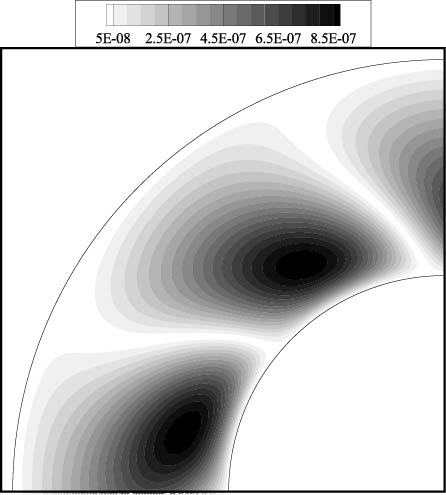}
}
\caption{(a) Pressure errors of the fluid cells near the boundary and (b) $\lambda$ error contours for the simulation of Taylor-Couette flow using the classical gas-kinetic BGK scheme, $\Delta x = d/75$.}
\end{figure}

First, the classical BGK scheme with a full expansion of the distribution function Eq~.\ref{eq:f_expansion} is employed to solve the flow field. The ${L_\infty }$ and ${L_2}$ error norms of the velocity $U_\theta$, pressure $p$ and $\lambda$ (related to the temperature) comparing with the analytical solution for different spatial resolutions are shown in Fig.~\ref{fig:test1_gksL}, in which the calculation of the error norms involves all of the fluid cells in the flow field. It is demonstrated that, for the velocity, the $L_\infty$-measure accuracy is around $2.50$ and $L_2$-measure accuracy is $2.62$, both are beyond the expected second-order accuracy, although in our method all interpolations and constructions are no more than second-order spacial accuracy. This may result from the so-called ``super-convergence'' of the classical BGK scheme \cite{Torrilhon2006Stability}. It's worth pointing out that although the convergence rate is faster, for the coarse grid the error of BGK scheme seems larger than that of the conventional NS solver. For the pressure, the spacial accuracy is $2.40$ in $L_2$ sense while only $1.45$ in $L_\infty$ sense, and the curve for $L_\infty$ error norm is somewhat zigzag. The pressure errors on the fluid cells near the inner cylinder are shown in Fig.~\ref{fig:test1_gks_pre}. The figure demonstrates that very high pressure error will occur on the fluid cell part of which lies on the solid side of the immersed boundary, which implies that the data from these cells may be not reliable. Another notable thing is that this problem is much lighter near the outer cylinder where the boundary is concave. The maximum pressure error of the fluid cell near the outer cylinder is one order of magnitude smaller than that near the inner cylinder. Thus, we think these singular fluid cells are due to the singularity of the ``ghost flow field'', i.e. one piece of flow field lies on the solid side of the immersed boundary (referred to as the ``ghost flow field'') may corresponding to several pieces of real flow field lie on the fluid side thus variables on the ghost flow field may have singular values, which is not suffered by a concave boundary. This is just similar to the singularity of the ghost cell discussed in Subsection \ref{subsec:ib}. However, these singular fluid cells seem not to spoil the order of the local accuracy as we exclude the fluid cells adjoining one or more solid cells from the calculation of the error norms and show the result in Fig.~\ref{fig:test1_gksLf}. It can be seen that, the accuracy of pressure achieves $2.05$ in $L_\infty$ sense and $2.68$ in $L_2$ sense. For $\lambda$, as shown in Fig.~\ref{fig:test1_gksL} or Fig.~\ref{fig:test1_gksLf}, the spacial accuracies in $L_\infty$ and $L_2$ measure are both straightly first-order. Furthermore, the $\lambda$ error contours (Fig.~\ref{fig:test1_gks_lambda}) show that the max $\lambda$ error occurs near the middle of the gap between the two cylinders, which implies that it is the BGK scheme but not the IB method which gives rise to the low accuracy of $\lambda$.

\begin{figure}
\centering
\subfigure[]{
\includegraphics[width=0.45\textwidth]{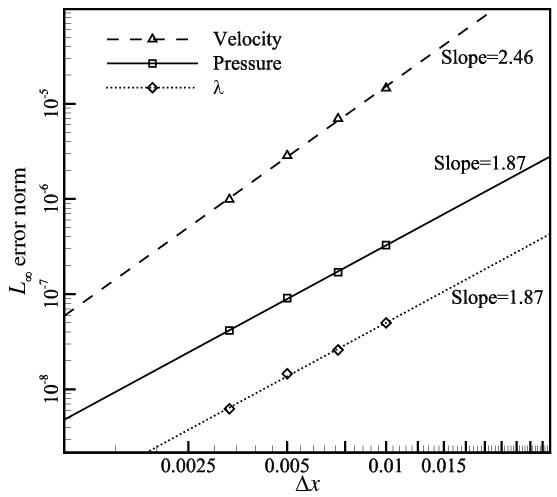}
}\hspace{0.05\textwidth}%
\subfigure[]{
\includegraphics[width=0.45\textwidth]{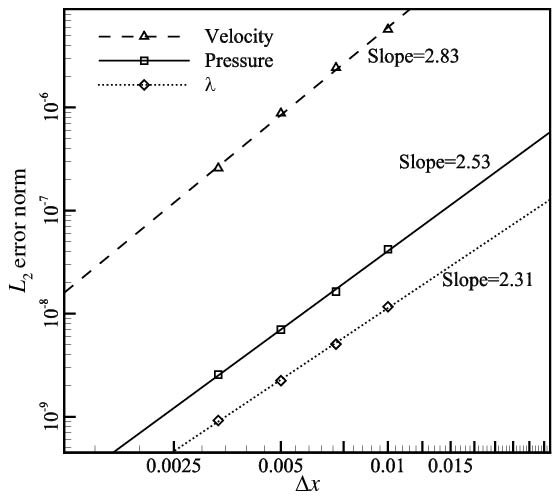}
}
\caption{\label{fig:test1_gnsL}(a) ${L_\infty }$ and (b) ${L_2 }$ error norms for the simulation of Taylor-Couette flow using the modified gas-kinetic BGK scheme. Norms are calculated from all fluid cells except those adjacent to solid cells.}
\end{figure}

Then, we reduce the full expansion of the distribution function (Eq~.\ref{eq:f_expansion}) to Eq.~\ref{eq:ns_f} and construct a simplified BGK solver to do some further investigation. It is a fact that a linear interpolation will give us a second-order estimation of the variable itself while a first-order estimation about the slope of the variable. Therefore, for Eq.~\ref{eq:ns_f}, if the linear reconstruction is employed just as what is done in the classical BGK scheme, the slope-associated terms, i.e. the terms multiplied by $\tau $, will be first-order accurate in space. In general, in the continuum limit, $\tau  \ll \Delta x$ and this first-order error will be covered up. However, when we use this simplified BGK scheme with linear reconstruction to do the above simulation of Taylor-Couette flow, we get a result similar to what is shown in Fig.~\ref{fig:test1_gksLf} and first-order accuracy of $\lambda$ is acquired. After that, we add compensatory terms to Eq.~\ref{eq:ns_f} to fix the first-order errors of the terms multiplied by $\tau $ and use this modified BGK scheme to do the same simulation. The $L_\infty$ and $L_2$ error norms for different spacial resolutions are shown in Fig.~\ref{fig:test1_gnsL}. It can be seen that, for the accuracy of velocity or pressure, there is no significant difference from the previous result (Fig.~\ref{fig:test1_gksLf}), but the spacial accuracy of $\lambda$ is increased to $1.87$ in $L_\infty$ sense and $2.31$ in $L_2$ sense. We have made an analysis for this and put forward the idea that it is the specificity of this test case which causes the low $\lambda$ accuracy when using the classical BGK scheme with linear reconstruction. In the Taylor-Couette flow, the heat convection is ignorable and the temperature of the fluid, which is related to $\lambda$, is dominated by the viscosity and heat conduction, which are only determined by the slope-associated terms (i.e. the nonequilibrium
part of the gas distribution function) in the full-expansion distribution function Eq~.\ref{eq:f_expansion} or the simplified distribution function Eq.~\ref{eq:ns_f}. Thus, if one wants to get second-order accuracy about the temperature (or $\lambda$) in such a heat-convection-ignorable test case, the slopes of the variables should be reconstructed in a second-order manner.

Finally, for the IB method, we can conclude that the local spacial accuracy (in $L_\infty$ sense, excluding the fluid cells adjacent to solid cells) is second-order and the global spacial accuracy (in $L_2$ sense) is also at least second-order.

\subsection{Supersonic flow around a stationary circular cylinder}
\begin{figure}
\centering
\subfigure[]{
\includegraphics[width=0.45\textwidth]{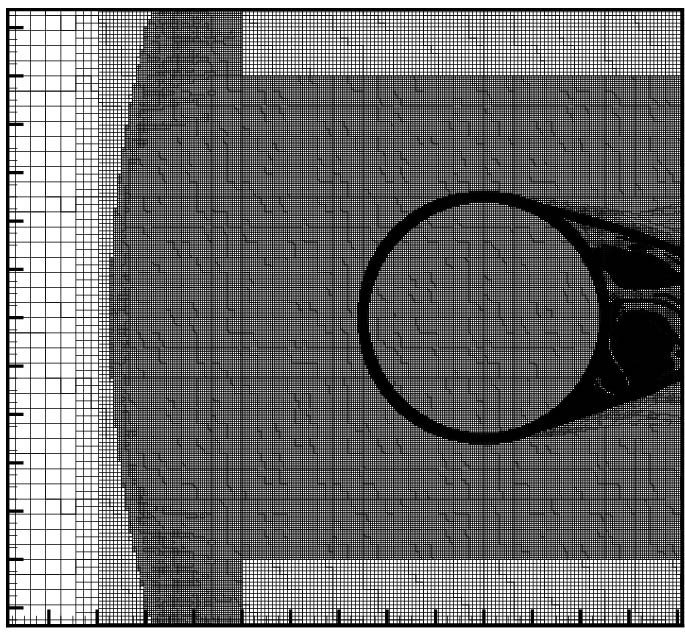}
}\hspace{0.05\textwidth}%
\subfigure[]{
\includegraphics[width=0.45\textwidth]{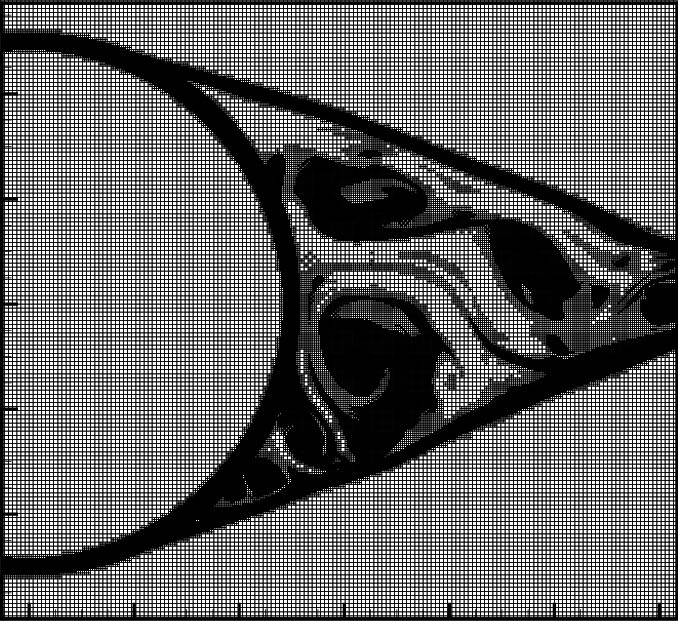}
}
\caption{\label{fig:test2_mesh}Instaneous nonuniform solution-adaptive Cartesian grid for the supersonic flow over a circular cylinder.}
\end{figure}

The supersonic turbulent flow past a stationary circular cylinder has been considered as a test case to show the performance of the present method on simulating highly compressible flow. The freestream Mach number is $\rm{Ma}=1.7$. The Reynolds number based on the freestream conditions and the cylinder diameter $d$ is $\rm{Re}=400000$. The Prandtl number is set to be $\rm{Pr}=0.71$. A rectangular computational domain with a size of $10d \times 30d$ is adopted where the cylinder is placed at $(2d,15d)$. In the far field, a coarse grid with the size $\Delta x = d/2$ is used. Approaching the cylinder, the grid is gradually refined and the finest grid is used around the immersed boundary. Besides, the solution-adaptive local grid refinement is employed and the grid will be refined at shear layers and shocks at every time step, as shown in Fig.~\ref{fig:test2_mesh}. The flow field is directly solved by the classical gas-kinetic BGK scheme without any turbulence-simulation technique, which makes the simulation very challenging. In view of the computational cost, three sizes of the near-wall grid, $\Delta x = d/256$, $d/512$ and $d/1024$ are considered, although even the finest size $d/1024$ seems not refined enough to resolve the turbulent flow.

\begin{figure}
\centering
\subfigure[\label{fig:test2_str}]{
\includegraphics[width=0.45\textwidth]{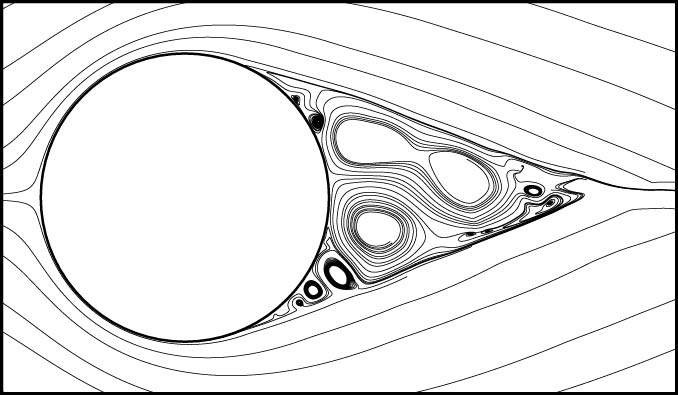}
}\hspace{0.05\textwidth}%
\subfigure[\label{fig:test2_vorc}]{
\includegraphics[width=0.45\textwidth]{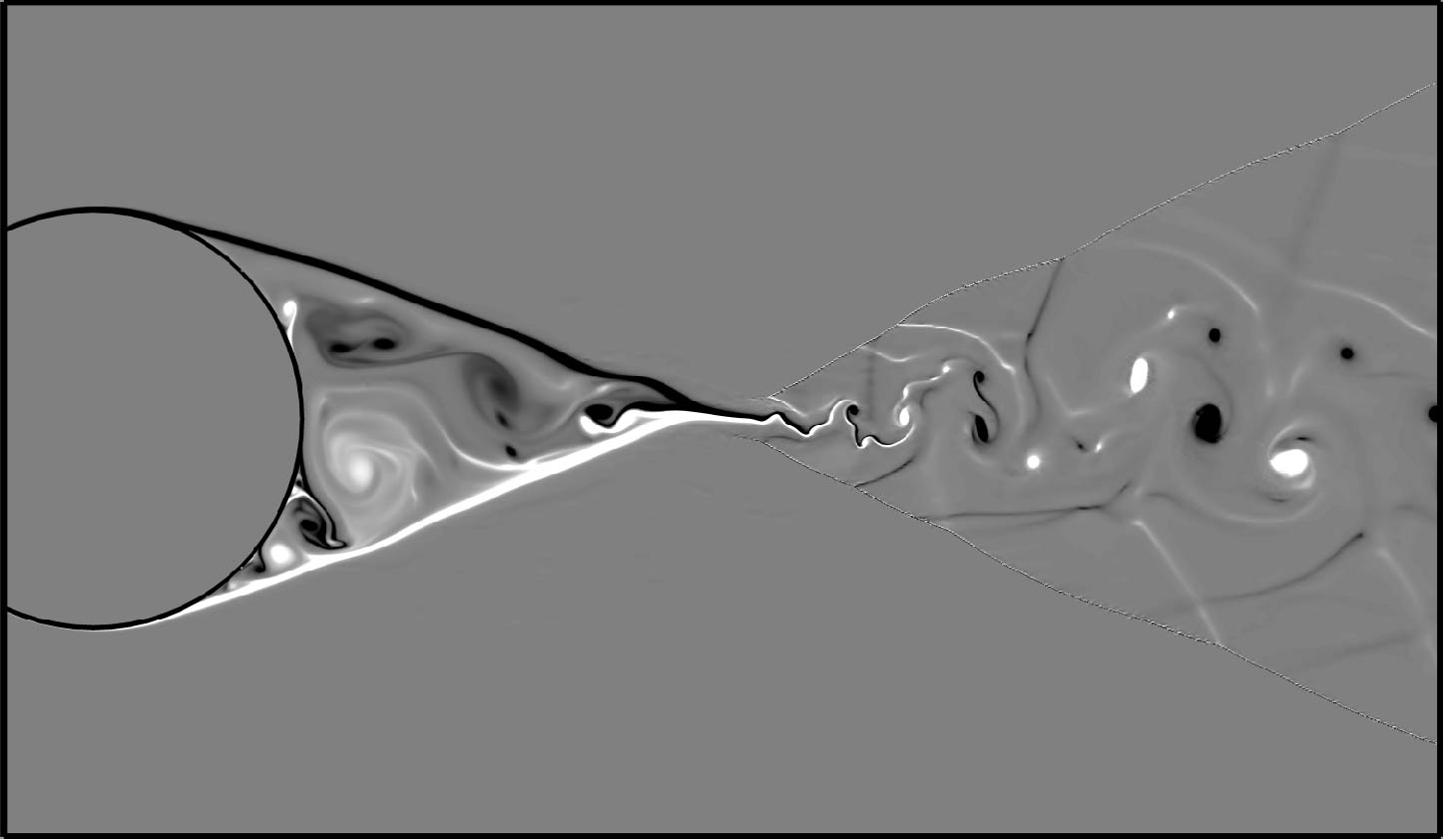}
}
\caption{\label{fig:test2_str_vorc}(a) Instaneous flow pattern and (b) the corresponding vorticity field for the supersonic flow over a circular cylinder, $\rm{Ma}=1.7$ and $\rm{Re}=400000$.}
\end{figure}

\begin{figure}
\centering
\includegraphics[width=0.75\textwidth]{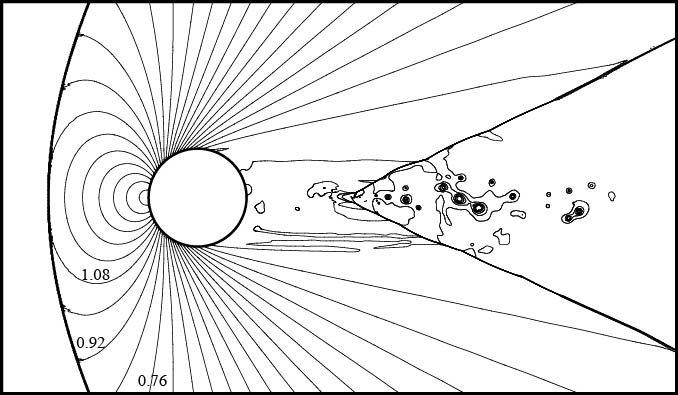}
\caption{\label{fig:test2_cpc}Instaneous pressure coefficient contours for the supersonic flow over a circular cylinder at $\rm{Ma}=1.7$ and $\rm{Re}=400000$, $\Delta {C_p}=0.08$.}
\end{figure}

\begin{figure}
\centering
\includegraphics[width=0.75\textwidth]{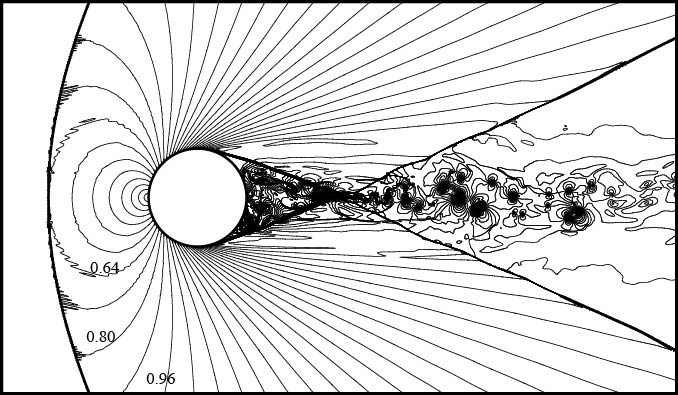}
\caption{\label{fig:test2_ma}Instaneous Mach number contours for the supersonic flow over a circular cylinder at $\rm{Ma}=1.7$ and $\rm{Re}=400000$, $\Delta {\rm{Ma}}=0.08$.}
\end{figure}

The instaneous flow pattern and the corresponding vorticity field, pressure coefficient contours and Mach number contours obtained from the finest grid are shown in Fig.~\ref{fig:test2_str_vorc}, Fig.~\ref{fig:test2_cpc} and Fig.~\ref{fig:test2_ma} respectively. Despite of the insufficient spatial resolution, some details of the flow can be observed. As what is displayed in the figures, a bow shock is formed ahead of the cylinder and there is a subsonic region behind the shock. The subsonic stream expands and accelerates along the surface of the cylinder, soon developing to the supersonic flow, separating at the rear of the the cylinder. Then a subsonic turbulent recirculation region forms behind the cylinder, enveloped by the separated supersonic flow. A strong shear layer can be observed between the subsonic recirculation region and the supersonic flow (Fig.~\ref{fig:test2_vorc}). Finally the supersonic streams from the upper and lower surfaces of the cylinder converge, forming two oblique shocks in the wake. The Mach number contours (Fig.~\ref{fig:test2_ma}) look very compatible with the published data in \cite{de2006immersed} and \cite{de2007immersed}.

\begin{figure}
\centering
\includegraphics[width=0.45\textwidth]{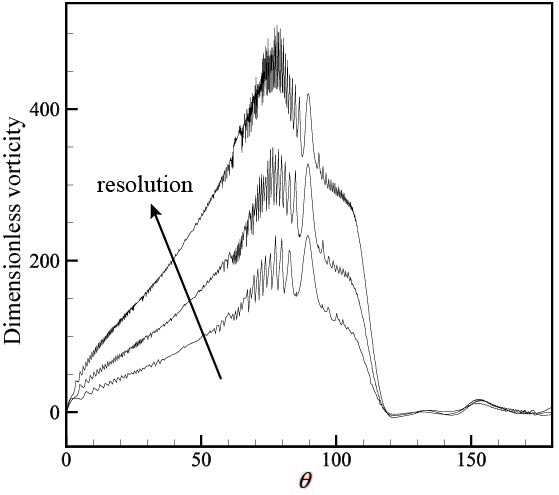}
\caption{\label{fig:test2_vor}Mean dimensionless vorticity distributions for the supersonic flow over a circular cylinder at $\rm{Ma}=1.7$ and $\rm{Re}=400000$, near-wall spatial resolutions $\Delta x = d/256, d/512$ and $d/1024$.}
\end{figure}

The mean vorticity distributions nondimensionalized by the freestream velocity and the diameter of the cylinder for different near-wall grid sizes are shown in Fig.~\ref{fig:test2_vor}. It can be seen that the vorticity curves are suffered from some oscillations. These oscillations are mainly due to the inadequate spatial resolution near the solid boundary and the way we obtain the vorticity, i.e. directly calculating the velocity gradient at the boundary-intercept point from the velocity of the corresponding image point, in which the velocity of the image point may be constructed oscillatory along the solid boundary if the grid is not refined enough. Despite of the oscillations, it can be observed that the vorticity increases significantly as the grid is refined, which implies that the solution may be still far from convergence at the finest grid. Besides, the locations of separation $\theta_{\rm{sep}}$ for different near-wall spatial resolutions determined from Fig.~\ref{fig:test2_vor} are given in Tab.~\ref{tab:test2_cdsep}, where the total drag coefficients are also listed. The present results are compared with the experimental result of Ignatova and Karimullin \cite{bashkin2002comparison} and the numerical results of De Palma et al.~\cite{de2006immersed} and de Tullio et al.~\cite{de2007immersed}. It can be found that for the drag coefficient $C_D$, the present results are consistent with the experimental result and also agree reasonably well with other numerical results. For the location of separation $\theta_{\rm{sep}}$, the present results appear non-convergent and deviate from most of the experiment and numerical results by around $6^\circ $ except the numerical result obtained by De Palma et al.~\cite{de2006immersed} from their coarse grid, which agrees well with the present $\theta_{\rm{sep}}$.

\begin{table}
\centering
\caption{\label{tab:test2_cdsep}Mean drag coefficient $C_D$ and mean location of separation ${\theta_{\rm{sep}}}$ for supersonic flow over a circular cylinder at $\rm{Ma} = 1.7$ and $\rm{Re} = 400000$.}
\begin{threeparttable}
\begin{tabular}{p{220pt} p{40pt}<{\centering} p{40pt}<{\centering}}
    \hline

    \hline
    \multicolumn{1}{c}{Work} & $C_D$ & ${\theta_{\rm{sep}}}$\\
     \hline
    Present ($\Delta x = d/256$) & 1.422 & $118.9^\circ $\\
    Present ($\Delta x = d/512$) & 1.436 & $118.4^\circ $\\
    Present ($\Delta x = d/1024$) & 1.446 & $118.6^\circ $\\
    Ignatova and Karimullin \cite{bashkin2002comparison} & 1.43\tnote{*} & $112^\circ $\\
    De Palma et al. (IB method, coarse) \cite{de2006immersed} & 1.38 & $118^\circ $\\
    De Palma et al. (body fitted, fine) \cite{de2006immersed} & 1.40 & $112^\circ $\\
    De Tullio et al.~\cite{de2007immersed} & 1.41 & $113^\circ $\\
    \hline

    \hline
\end{tabular}
\begin{tablenotes}
\item[*] The experimental pressure drag coefficient in \cite{bashkin2002comparison} is taken as the total drag coefficient approximately.
\end{tablenotes}
\end{threeparttable}
\end{table}

\begin{figure}
\centering
\subfigure[\label{fig:test2_cp_prt}]{
\includegraphics[width=0.45\textwidth]{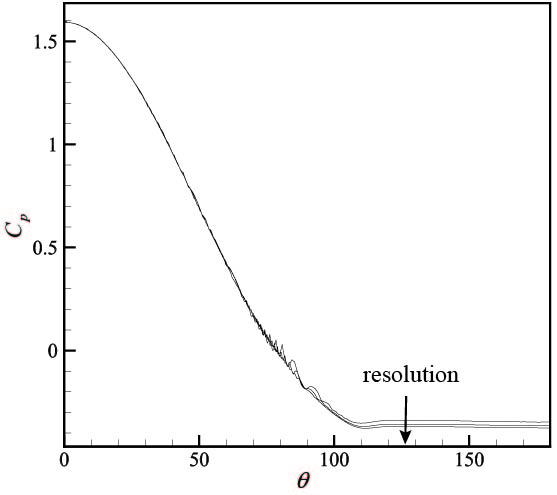}
}\hspace{0.05\textwidth}%
\subfigure[\label{fig:test2_cp_cmp}]{
\includegraphics[width=0.45\textwidth]{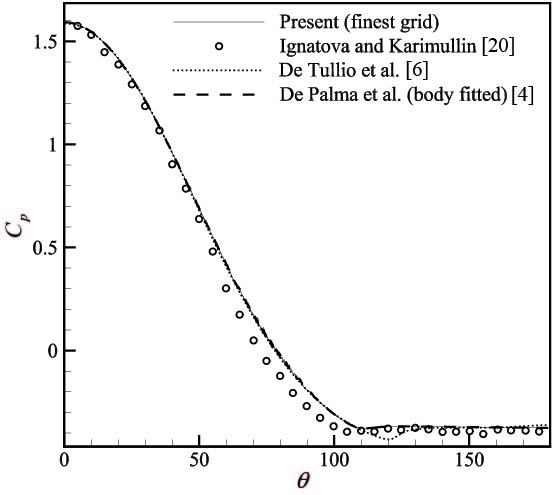}
}
\caption{\label{fig:test2_cp}Mean pressure coefficient distributions for the supersonic flow over a circular cylinder at $\rm{Ma}=1.7$ and $\rm{Re}=400000$, (a) the present results for different near-wall spatial resolutions $\Delta x = d/256, d/512, d/1024$ and (b) the comparison among the present and other authors' experimental(circles)/numerical results.}
\end{figure}

The mean pressure coefficient distributions for different near-wall spatial resolutions are plotted and compared with other authors' results in Fig.~\ref{fig:test2_cp}. Fig.~\ref{fig:test2_cp_prt} shows that as the grid is refined, the oscillation of the pressure coefficient distribution is suppressed. This oscillation about the pressure can be also seen in the result of De Palma et al.~\cite{de2006immersed} obtained by their IB method, which seems like a common problem for the class of IB methods. Besides, it can be observed that as the spatial resolution increases, the pressure at the rear of the cylinder will decrease, leading to the increase in total drag (Tab.~\ref{tab:test2_cdsep}). In Fig.~\ref{fig:test2_cp_cmp}, the present result and the numerical result of De Palma et al.~\cite{de2006immersed} obtained from their body-fitted mesh almost overlap, showing great consistence, and they are also in good agreement with the numerical result of de Tullio et al.~\cite{de2007immersed}. All three numerical results agree with the experimental data reasonably well. The most significant difference between the experimental and numerical results is that at $\theta  > 50^\circ $ up to the separation point the experimental curve lies below the numerical curves. This difference can be attributed to the spatial effect in the experiment \cite{bashkin2002comparison}, which is ignored in the two-dimensional numerical simulation.

In general, for this test case, without any turbulence-simulation technique and in the insufficient spatial resolution, the present method fails to give a smooth vorticity distribution and a very accurate separation point. However, there are still a lot of flow characteristics can be resolved correctly. With proper turbulence model or LES technique, we are confident that the present method can yield a satisfactory result for this supersonic turbulent problem.

\subsection{Incompressible flow around an oscillating circular cylinder}
The test case of a circular cylinder oscillating incompressibly in stationary fluid has been conducted to demonstrate the ability of the present method in manipulating the moving boundary. In this test case, a circular cylinder oscillates sinusoidally in the horizontal direction with the equation of motion which can be expressed as
\begin{equation}
x(t) =  - A\sin (2\pi ft),
\end{equation}
where $x$ is the horizontal displacement of the cylinder, $A$ is the amplitude and $f$ is the frequency. Two key parameters dominating the pattern of this flow are the Reynolds number Re and the Keulegan-Carpenter number KC, which are defined as
\begin{equation}
\begin{aligned}
{\mathop{\rm Re}\nolimits}  &= \frac{{\rho {U_{\max }}d}}{\mu },\\
{\rm{KC}} &= \frac{{{U_{\max }}}}{{fd}},
\end{aligned}
\end{equation}
where $d$ is the diameter of the cylinder, $U_{\max}$ is the maximum horizontal velocity of the cylinder, $\rho$ is the fluid density and $\mu$ is the fluid viscosity. In our simulation, the condition $\rm{Re} = 100, \rm{KC} = 5$ is considered and the maximum velocity $U_{\max}$ is set to be equal to a Mach number $\rm{Ma}=0.1$ to approximate the incompressible flow. According to Tatsuno and Bearman \cite{tatsuno1990osc} and our previous work \cite{Yuan2015An}, for this parameter set the flow is two-dimensional, stable and symmetric. When the cylinder moves through the equilibrium position, the maximum velocity is attained and two counter-rotating vortices with the same magnitude of strength are formed behind the cylinder. Then the cylinder will reach the maximum displacement and move back, hitting at the pair of vortices which will be split and finally reverse during the backward motion of the cylinder.

\begin{figure}
\centering
\includegraphics[width=0.45\textwidth]{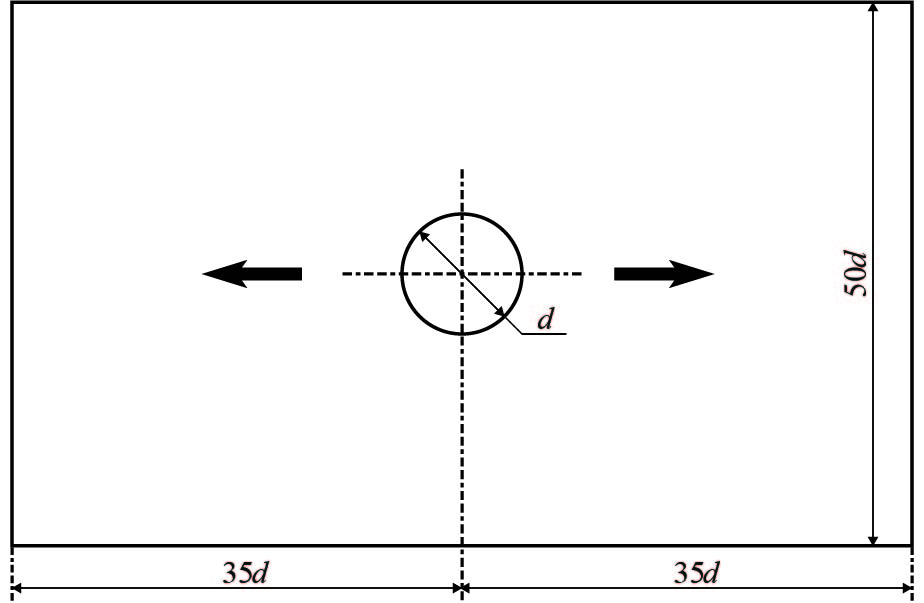}
\caption{\label{fig:test3_oscdomain} Schematic for the computational domain of the flow around an oscillating circular cylinder.}
\end{figure}

\begin{figure}
\centering
\includegraphics[width=0.45\textwidth]{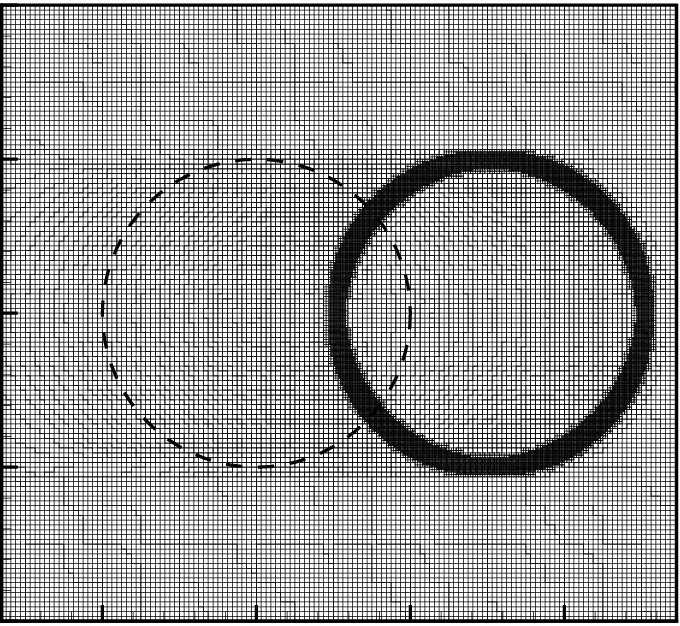}
\caption{\label{fig:test3_oscmesh} Mesh for the flow around an oscillating circular cylinder at phase position $288^\circ $. The dashed circle represents the equilibrium position.}
\end{figure}

As shown in Fig.~\ref{fig:test3_oscdomain}, a rectangular computational domain with the size $70d \times 50d$ is adopted in this test case. The equilibrium position of the cylinder's sinusoidal motion is set at $(35d,25d)$. A nonuniform Cartesian grid, as described in Subsection \ref{subsec:grid}, is adopted with a large mesh size equal to $1d$ in the far field and a refined size $d/256$ (refined enough \cite{Yuan2015An}) near the cylinder. The refined region around the immersed boundary will move along with the motion of the cylinder, as what is displayed in Fig.~\ref{fig:test3_oscmesh}. The classical gas-kinetic BGK scheme is employed in the simulation.

\begin{figure}
\centering
\subfigure[]{
\includegraphics[width=0.45\textwidth]{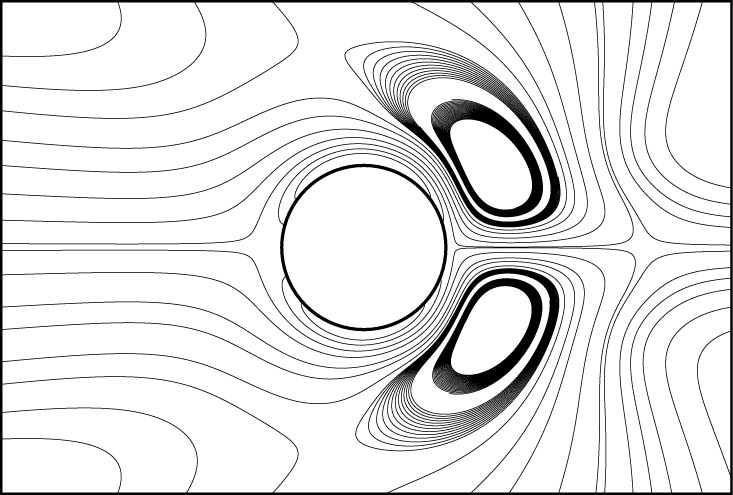}
}\hspace{0.05\textwidth}%
\subfigure[]{
\includegraphics[width=0.45\textwidth]{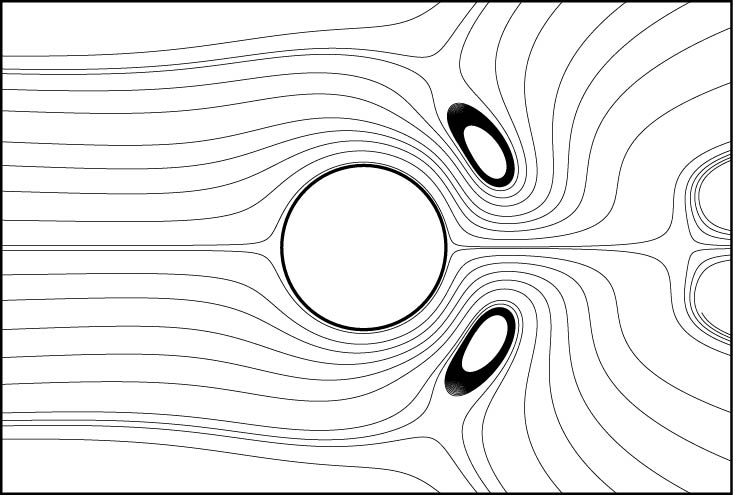}
}
\caption{\label{fig:test3_096st} Instantaneous flow patterns for flow around an oscillating circular cylinder at phase position $96^\circ $ in (a) reference frame which is fixed in the space and (b) reference frame which is fixed on the cylinder.}
\end{figure}

The instantaneous flow pattern at phase angle $96^\circ $ is shown in Fig.~\ref{fig:test3_096st}. At this moment the cylinder has just reached the maximum negative displacement and is moving back to its equilibrium position, hitting at the two vortices formed behind it. It can be seen that the stream line fits the boundary completely and no flow penetration can be observed.

\begin{figure}
\centering
\subfigure[]{
\includegraphics[width=0.45\textwidth]{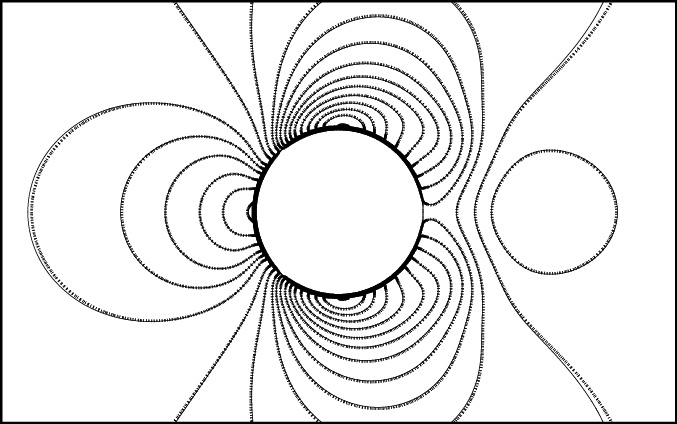}
\hspace{0.05\textwidth}
\includegraphics[width=0.45\textwidth]{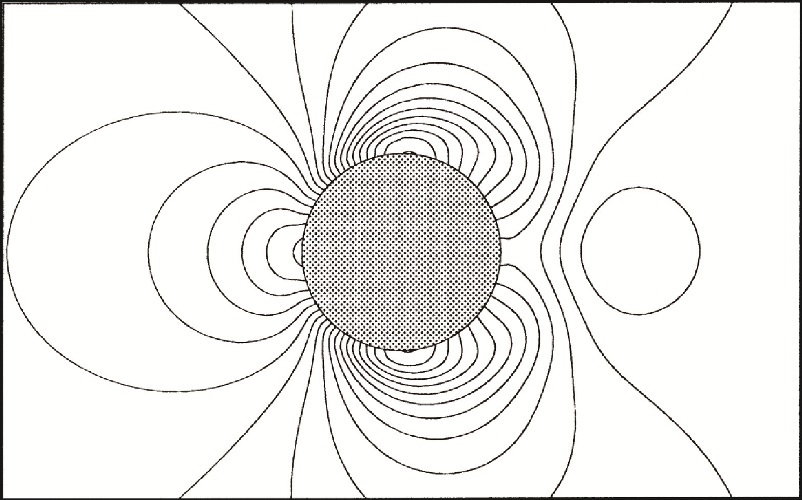}
}
\subfigure[]{
\includegraphics[width=0.45\textwidth]{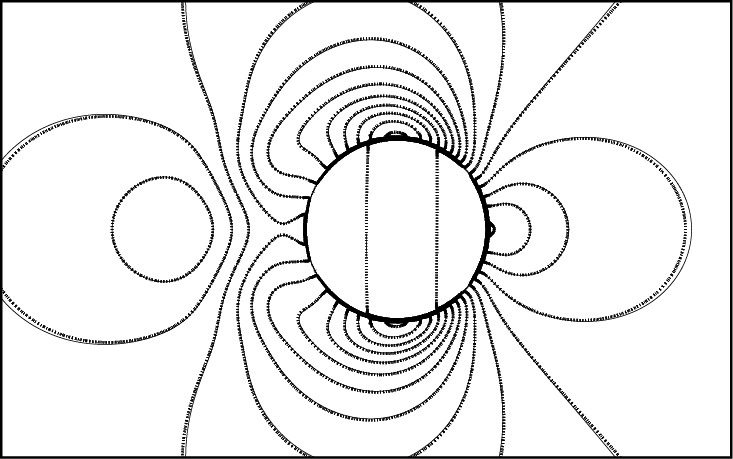}
\hspace{0.05\textwidth}
\includegraphics[width=0.45\textwidth]{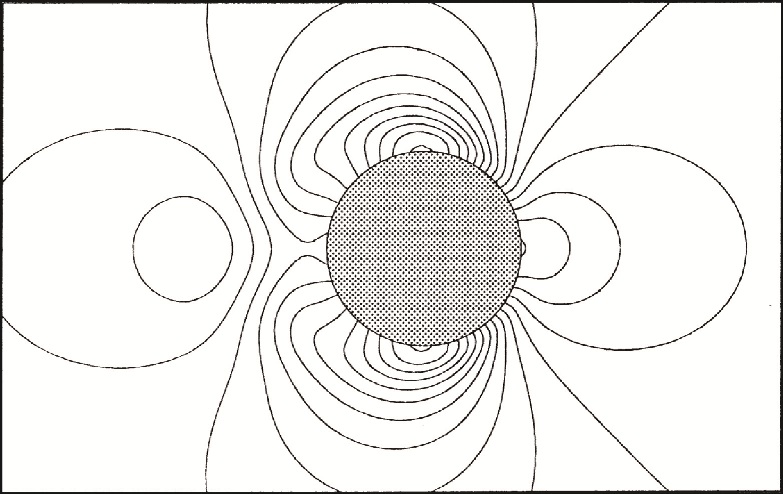}
}
\caption{\label{fig:test3_oscpre} Pressure isolines for flow around an oscillating circular cylinder at phase positions (a) $0^\circ $ and (b) $192^\circ $, the solid lines in the left figures are the present results, the dotted lines in the left figures are the results in Yuan et al.~\cite{Yuan2015An}, the right figures are the results of D\"utsch et al.~\cite{dutsch1998low}.}
\end{figure}

\begin{figure}
\centering
\subfigure[]{
\includegraphics[width=0.45\textwidth]{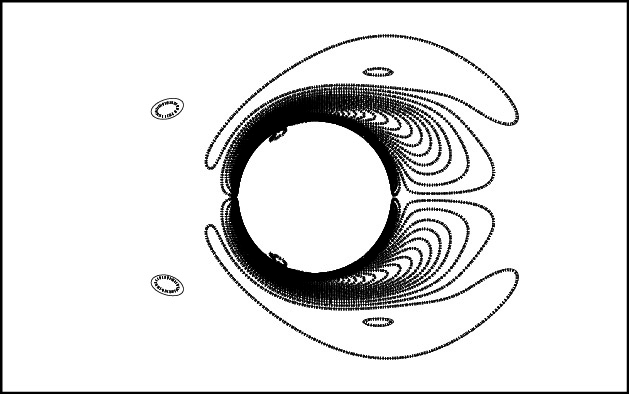}
\hspace{0.05\textwidth}
\includegraphics[width=0.45\textwidth]{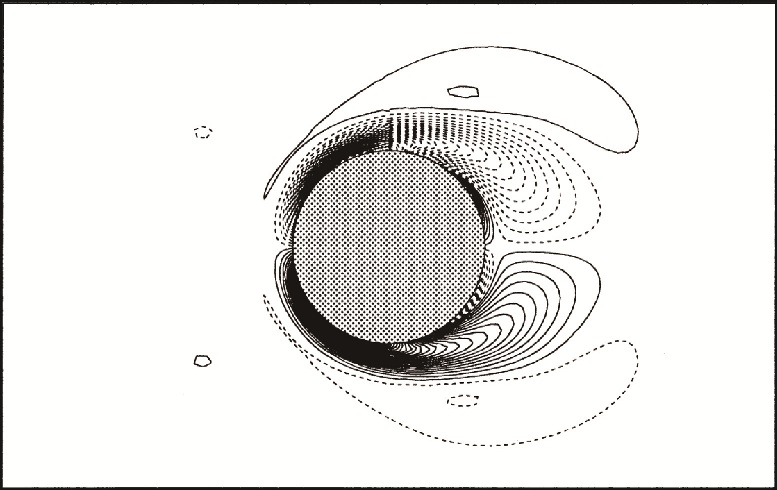}
}
\subfigure[]{
\includegraphics[width=0.45\textwidth]{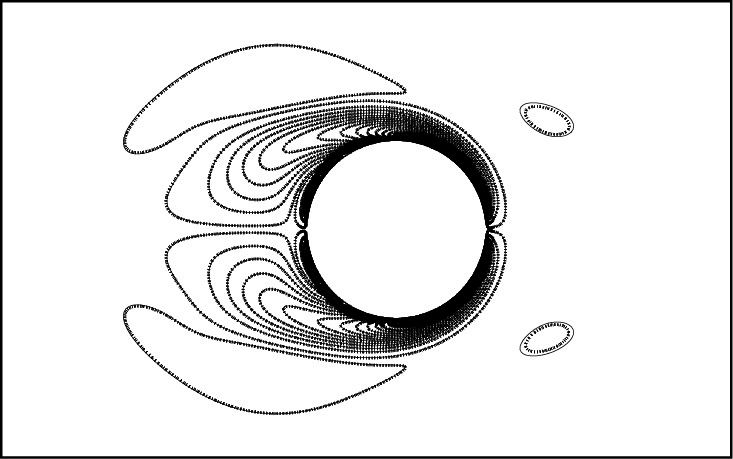}
\hspace{0.05\textwidth}
\includegraphics[width=0.45\textwidth]{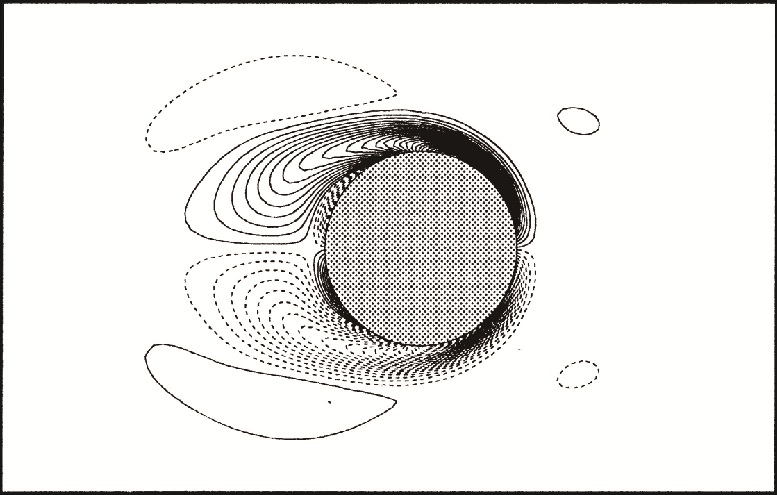}
}
\caption{\label{fig:test3_oscvor} Vorticity isolines for flow around an oscillating circular cylinder at phase positions (a) $0^\circ $ and (b) $192^\circ $, the solid lines in the left figures are the present results, the dotted lines in the left figures are the results in Yuan et al.~\cite{Yuan2015An}, the right figures are the results of D\"utsch et al.~\cite{dutsch1998low}.}
\end{figure}

The instantaneous pressure and vorticity isolines at phase positions $0^\circ $ and $192^\circ $ are shown in Fig.~\ref{fig:test3_oscpre} and Fig.~\ref{fig:test3_oscvor} respectively. In the left figures the present result is compared quantitatively with the result from Yuan et al.~\cite{Yuan2015An} (our previous work), where isolines from two results share the identical variable values. As is demonstrated in the left figures, the two results coincide with each other perfectly near the cylinder while minor mismatches of the isolines exist in the area a little further. These mismatches are attributable to the insufficient computing time in Yuan et al.~\cite{Yuan2015An}, which will make the flow non-fully developed in the far field. In the right figures of Fig.~\ref{fig:test3_oscpre} and Fig.~\ref{fig:test3_oscvor}, the numerical result of D\"utsch et al.~\cite{dutsch1998low} is exhibited (variable values of the isolines may be different from those in the left figures) and we can get some qualitative comparison between these three results. Generally speaking, the three results agree reasonably well.

\begin{figure}
\centering
\subfigure[]{
\includegraphics[width=0.45\textwidth]{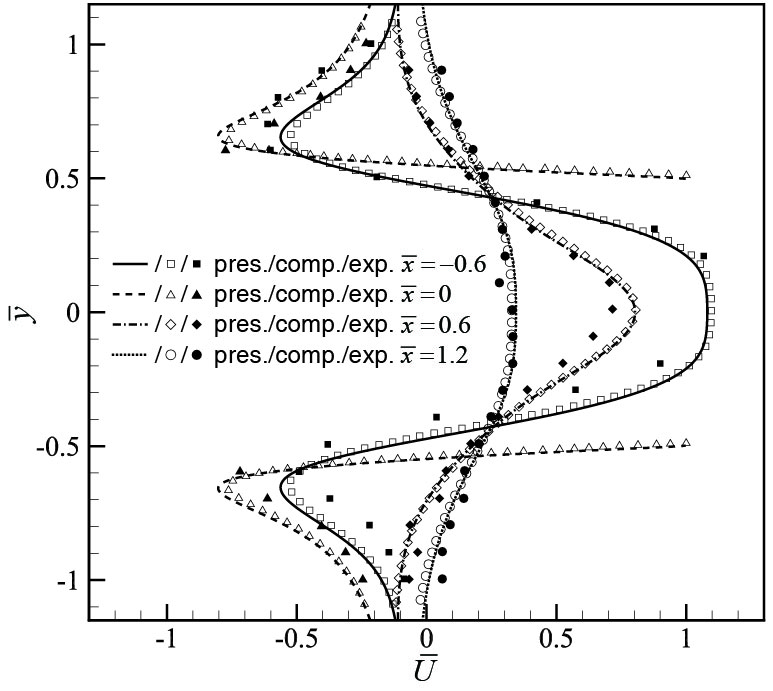}
\hspace{0.05\textwidth}
\includegraphics[width=0.45\textwidth]{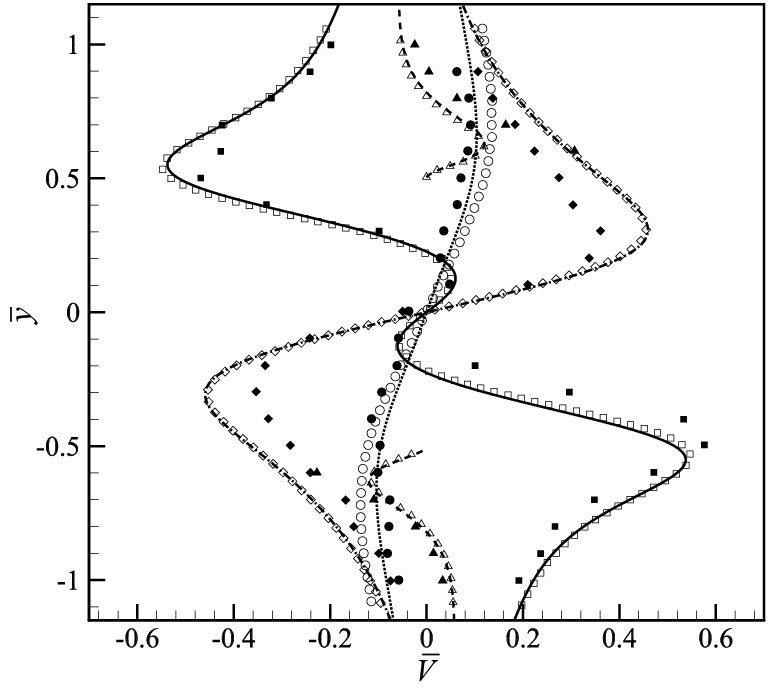}
}
\subfigure[]{
\includegraphics[width=0.45\textwidth]{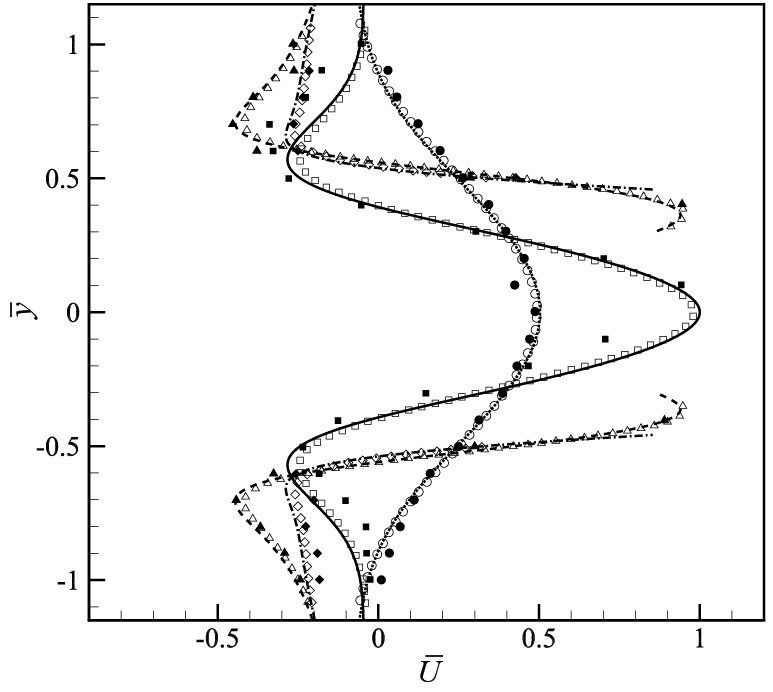}
\hspace{0.05\textwidth}
\includegraphics[width=0.45\textwidth]{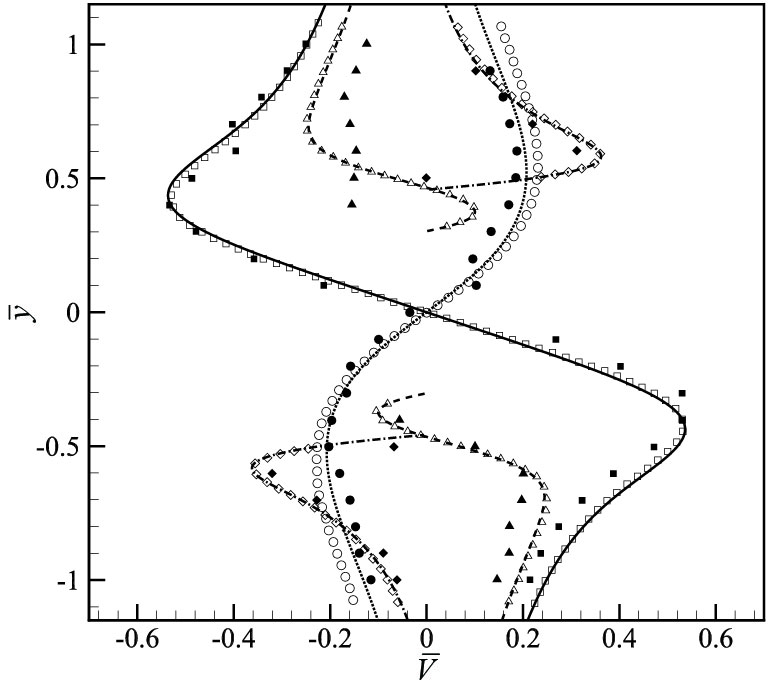}
}\\
\subfigure[]{
\includegraphics[width=0.45\textwidth]{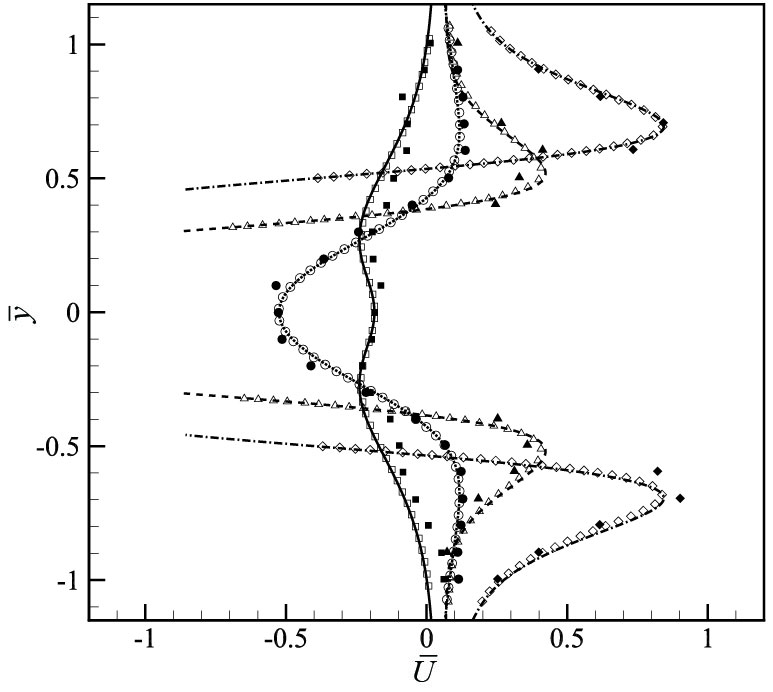}
\hspace{0.05\textwidth}
\includegraphics[width=0.45\textwidth]{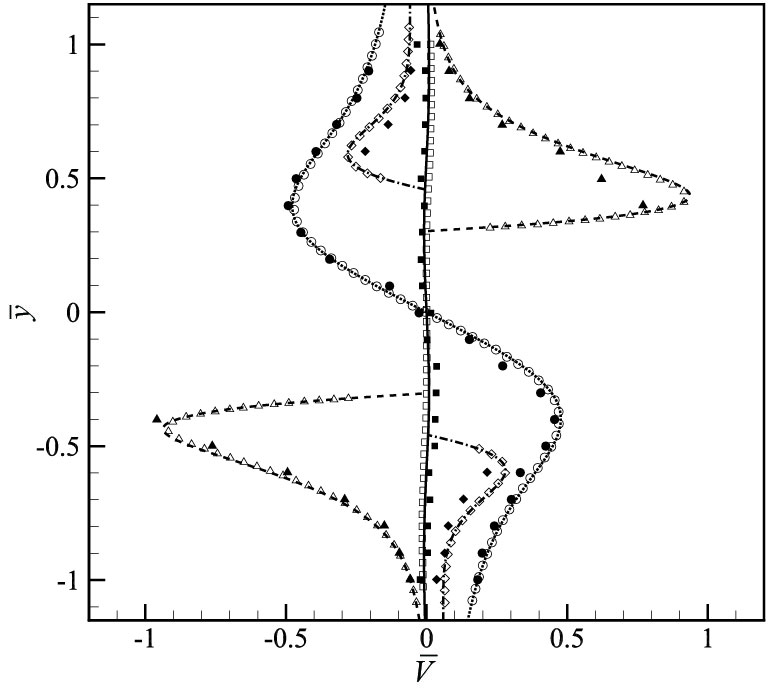}
}
\caption{\label{fig:test3_oscprf} Velocity profiles of the flow around an oscillating circular cylinder at four vertical cross sections $\bar x =  - 0.6,0,0.6,1.2$ for phase positions (a) ${180^ \circ }$, (b) ${210^ \circ }$ and (c) ${330^ \circ }$. The lines are the present results and the symbols are the computational/experimental results of D\"utsch et al.~\cite{dutsch1998low}.}
\end{figure}

The velocity profiles near the cylinder at four vertical cross sections $\bar x =  - 0.6,0,0.6,1.2$ for three phase positions ${180^ \circ }$, ${210^ \circ }$ and ${330^ \circ }$ are shown in Fig.~\ref{fig:test3_oscprf}. Here, $\bar x,\bar y,{\bar U}$ and ${\bar V}$ are defined as
\begin{equation}
\bar x = \frac{x}{d},\;\bar y = \frac{y}{d},\;{\bar U} = \frac{{{U}}}{{{U_{\max }}}},\;{\bar V} = \frac{{{V}}}{{{U_{\max }}}},
\end{equation}
where $x, y$ are the coordinates relative to the equilibrium position of the sinusoidal oscillation, $U, V$ are the velocity components in the horizontal and vertical direction. The present result is compared with the numerical and experimental results of D\"utsch et al.~\cite{dutsch1998low}. Generally good agreement is obtained among the three sets of results.

\begin{figure}
\centering
\includegraphics[width=0.45\textwidth]{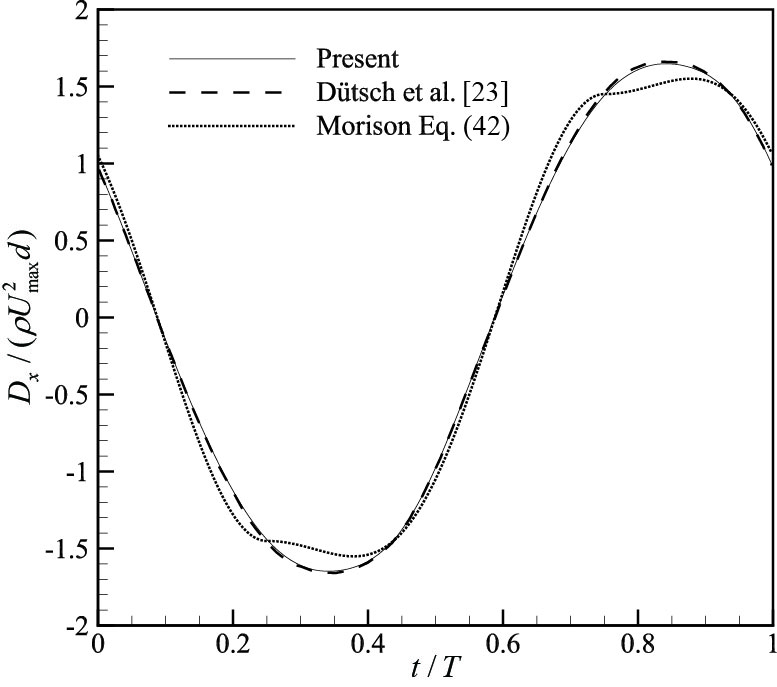}
\caption{\label{fig:test3_oscDx} Dimensionless inline force during an oscillating period $T$ for flow around an oscillating circular cylinder.}
\end{figure}

According to Morison et al.'s semi-empirical equation \cite{morison1950force}, which is widely used in estimating the inline force on the body in oscillatory flow, the time-dependent inline force $D_x$ on the cylinder oscillating in a stationary fluid can be expressed as
\begin{equation}\label{eq:Morison}
{D_x} =  - \frac{1}{2}\rho d{c_d}\dot x\left| {\dot x} \right| - \frac{1}{4}\pi \rho {d^2}{c_i}\ddot x,
\end{equation}
where $x$ is the displacement of the cylinder, $c_d$ and $c_i$ are the drag coefficient and the added mass coefficient. $c_d$ and $c_i$ are two empirical hydrodynamic coefficients and can be obtained by various methods such as Fourier analysis or least-squares fitting after we have got the time-dependent inline force $D_x$ from experiment or numerical simulation. In our numerical simulation, the inline force $D_x$ is obtained from Eq.~\ref{eq:Dx} and the values of $c_d$ and $c_i$ fitted to this force are shown and compared with other authors' numerical results in Tab.~\ref{tab:test3_cdci}. The present result agrees well with other three sets of results. The inline force calculated from the Morison equation Eq.~\ref{eq:Morison} with the present $c_d$ and $c_i$ is shown in Fig.~\ref{fig:test3_oscDx} and compared with the forces obtained directly from the numerical simulations of us and D\"utsch et al.~\cite{dutsch1998low}. It can be seen that the force calculated from the Morison equation is grossly consistent with the two numerical results, which agree with each other excellently.

\begin{table}
\centering
\caption{\label{tab:test3_cdci}Drag coefficient $c_d$ and added mass coefficient $c_i$ for flow over an oscillating circular cylinder in a stationary fluid.}
\begin{threeparttable}
\begin{tabular}{p{105pt} p{80pt}<{\centering} p{80pt}<{\centering}}
    \hline

    \hline
    \multicolumn{1}{c}{Work} & $c_d$ & $c_i$ \\
     \hline
    Present & $2.10$ & $1.47$ \\
    Yuan et al.~\cite{Yuan2015An} & $2.10$ & $1.45$ \\
    D\"utsch et al.~\cite{dutsch1998low}\tnote{*} & $2.09$ & $1.45$ \\
    Uzuno{\u{g}}lu et al.~\cite{uzunouglu2001low}\tnote{*} & $2.10$ & $1.45$ \\
    \hline

    \hline
\end{tabular}
\begin{tablenotes}
\item[*] Data are based on the finest meshes in \cite{dutsch1998low} and \cite{uzunouglu2001low}.
\end{tablenotes}
\end{threeparttable}
\end{table}

\subsection{Compressible flow around a moving airfoil}
\begin{figure}
\centering
\includegraphics[width=0.45\textwidth]{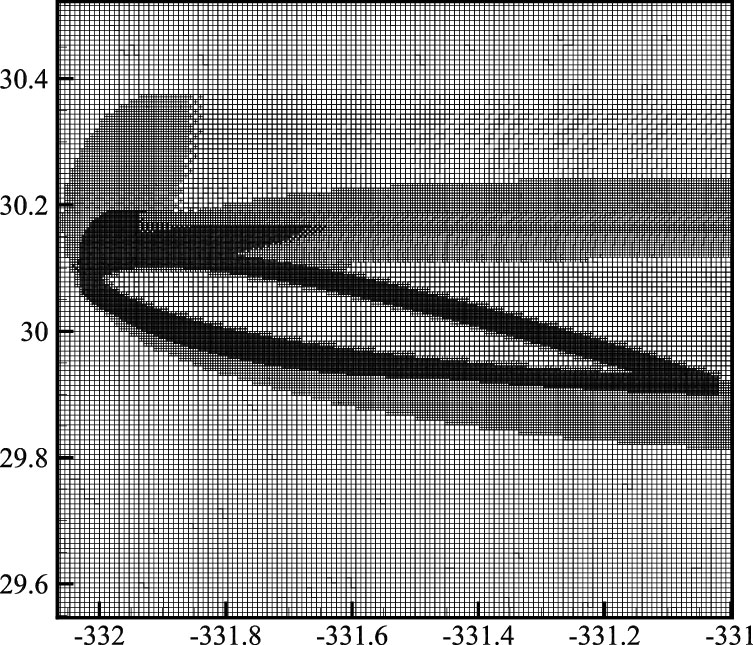}
\caption{\label{fig:test4_mesh} The nonuniform Cartesian grid used for the flow around a moving NACA0012 airfoil.}
\end{figure}

To further verify the ability of the present method in handling compressible moving- boundary problem, the test case of an airfoil passing through a compressible viscous fluid is conducted. In this simulation, a NACA0012 airfoil at an angle of attack $\alpha  = 10^\circ $ is moving at a Mach number ${\rm{Ma}} = 0.8$ in a stationary fluid. The Reynolds number based on the chord length $c$ and the freestream condition is ${\mathop{\rm Re}\nolimits}  = 500$. The airfoil is set to be insulated and the Prandtl number is $\Pr  = 0.71$. A rectangle computational domain with a large size $70c \times 60c$ is adopted, which will move along with the motion of the airfoil so that the airfoil will be always placed at the center of the computational domain. In the far field a coarse grid with a size of $1c$ is used while near the airfoil the grid is gradually refined to $c/512$, as shown in Fig.~\ref{fig:test4_mesh}. Also, the refined region will move with the airfoil. The classical gas-kinetic BGK scheme is employed in the simulation. Furthermore, as a control, the flow around a stationary NACA0012 airfoil with the same conditions, computational domain and grid is investigated.

\begin{figure}
\centering
\subfigure[]{
\includegraphics[width=0.45\textwidth]{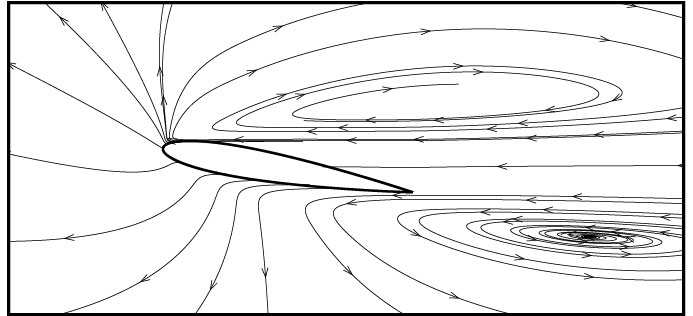}
}\hspace{0.05\textwidth}%
\subfigure[\label{fig:test4_str1}]{
\includegraphics[width=0.45\textwidth]{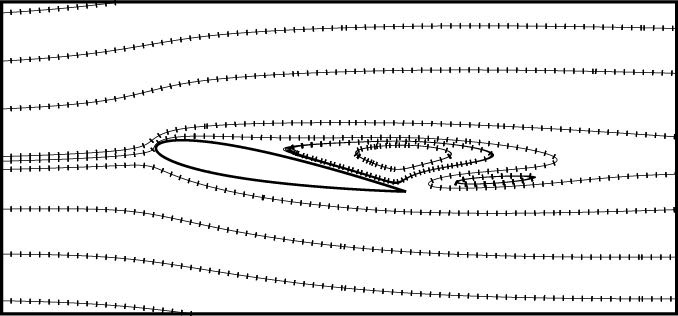}
}
\caption{\label{fig:test4_str} Streamlines for flow around a moving NACA0012 airfoil at $\rm{Ma} = 0.8$, $\rm{Re} = 500$ and $\alpha  = 10^\circ $ in (a) reference frame which is fixed in the space and (b) reference frame which is fixed on the airfoil. The solid lines in the figure (b) are results of flow around a moving airfoil while the dotted lines are results of flow around a stationary airfoil.}
\end{figure}

\begin{figure}
\centering
\subfigure[]{
\includegraphics[width=0.45\textwidth]{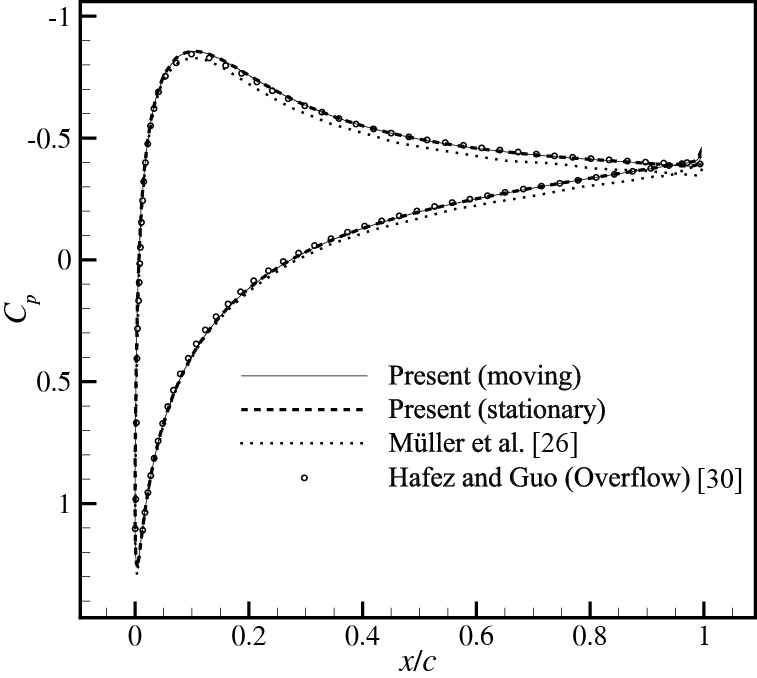}
}\hspace{0.05\textwidth}%
\subfigure[]{
\includegraphics[width=0.45\textwidth]{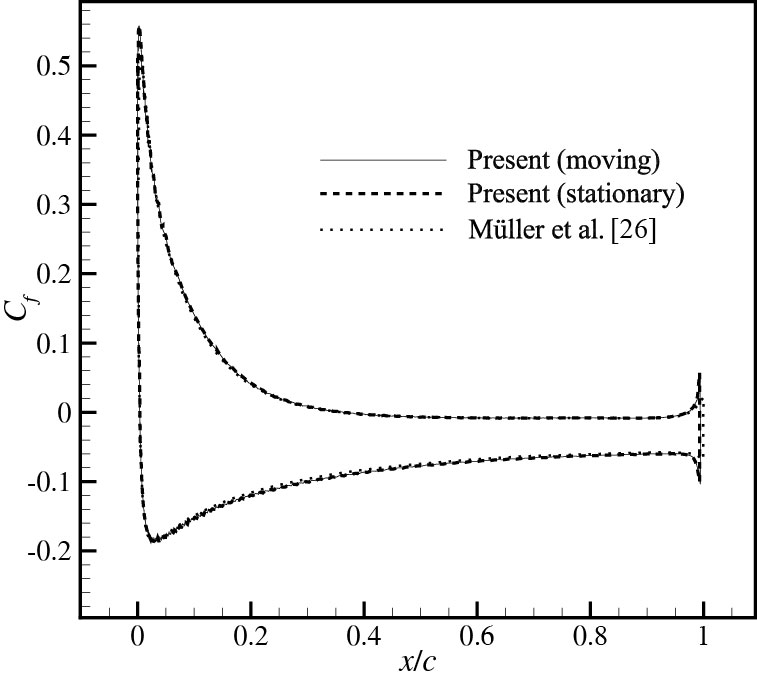}
}
\caption{\label{fig:test4_cpcf} (a) Pressure coefficient distributions and (b) skin friction coefficient distributions for flow around a NACA0012 airfoil at $\rm{Ma} = 0.8$, $\rm{Re} = 500$ and $\alpha  = 10^\circ $.}
\end{figure}

Fig.~\ref{fig:test4_str} shows the stream lines around the airfoil in the reference frame which is fixed in the space and the reference frame which is fixed on the airfoil. In Fig.~\ref{fig:test4_str1}, it can be seen that the two sets of flow patterns, which are based on the moving-airfoil simulation and the stationary-airfoil simulation respectively, are entirely coincident and fit the solid boundary perfectly. It is demonstrated that the flow separates at some place $x_{\rm{sep}}$ from the upper surface of the airfoil and a large recirculation region forms behind the airfoil. The location of the separation $x_{\rm{sep}}$, which is determined from the distribution of the skin friction coefficient $C_f$ (Fig.~\ref{fig:test4_cpcf}), the drag coefficient $C_D$ and the lift coefficient $C_L$, for which the drag and lift forces can be obtained by Eq.~\ref{eq:Dx}, are presented in Tab.~\ref{tab:test4_cdclsep}. It shows that the two sets of present results agree completely and they are also consistent with the results from other authors.

\begin{table}
\centering
\caption{\label{tab:test4_cdclsep}Drag coefficient $C_D$, lift coefficient $C_L$ and location of separation point ${x_{{\rm{sep}}}}$ for flow over a NACA0012 airfoil at $\rm{Ma} = 0.8$, $\rm{Re} = 500$ and $\alpha  = 10^\circ $.}
\begin{tabular}{p{140pt} p{40pt}<{\centering} p{40pt}<{\centering} p{40pt}<{\centering}}
    \hline

    \hline
    \multicolumn{1}{c}{Work} & $C_D$ & $C_L$ & ${x_{{\rm{sep}}}}/c$\\
     \hline
    Present (moving) & 0.2734 & 0.4347 & 0.363\\
    Present (stationary) & 0.2734 & 0.4347 & 0.363\\
    M{\"u}ller et al.~\cite{muller1987implicit} & 0.2632 & 0.4199 & 0.371\\
    Cambier \cite{cambier1987computation} & 0.2656 & 0.4342 & 0.36\\
    Kordulla \cite{kordulla1987using} & 0.2845 & 0.4261 & 0.362\\
    Paillere and Deconinck \cite{paillere1995multidimensional} & 0.2728 & 0.4530 & 0.383\\
    \hline

    \hline
\end{tabular}
\end{table}

\begin{figure}
\centering
\subfigure[]{
\includegraphics[width=0.45\textwidth]{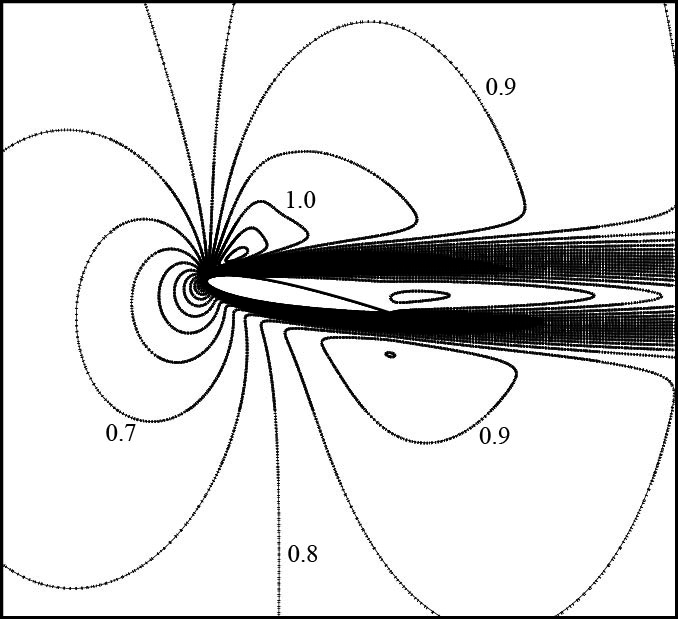}
}\hspace{0.05\textwidth}%
\subfigure[]{
\includegraphics[width=0.45\textwidth]{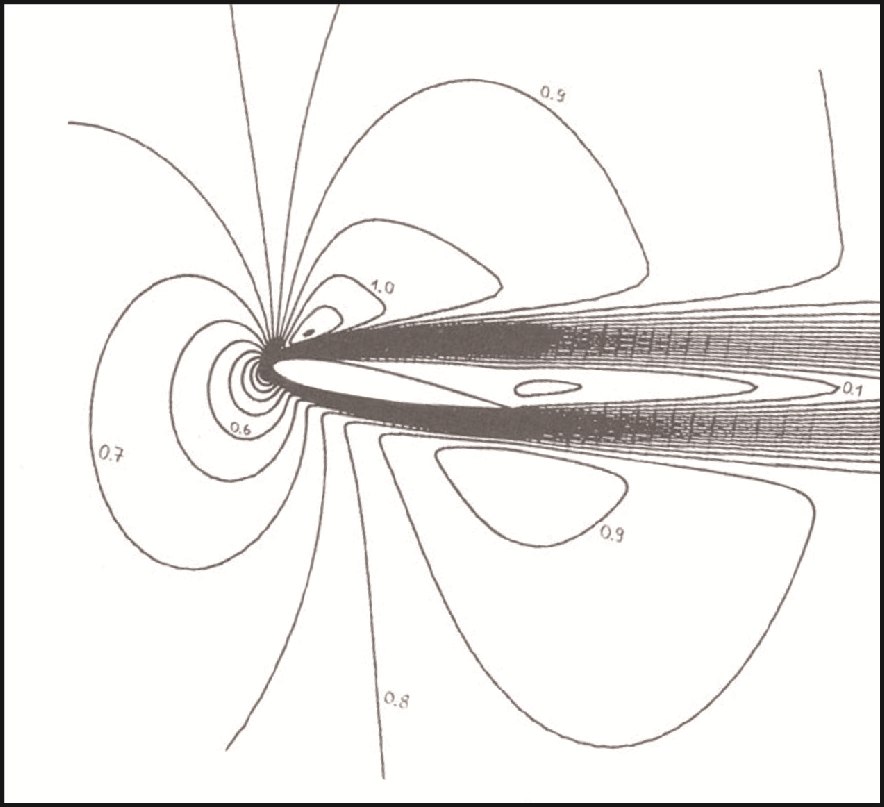}
}
\caption{\label{fig:test4_ma} Mach number contours for flow around a NACA0012 airfoil at $\rm{Ma} = 0.8$, $\rm{Re} = 500$ and $\alpha  = 10^\circ $, (a) the present result (solid lines are based on the moving airfoil, dotted lines are based on the stationary airfoil) and (b) the result of M{\"u}ller et al.~\cite{muller1987implicit}, $\Delta {\rm{Ma}} = 0.05$.}
\end{figure}

\begin{figure}
\centering
\subfigure[]{
\includegraphics[width=0.45\textwidth]{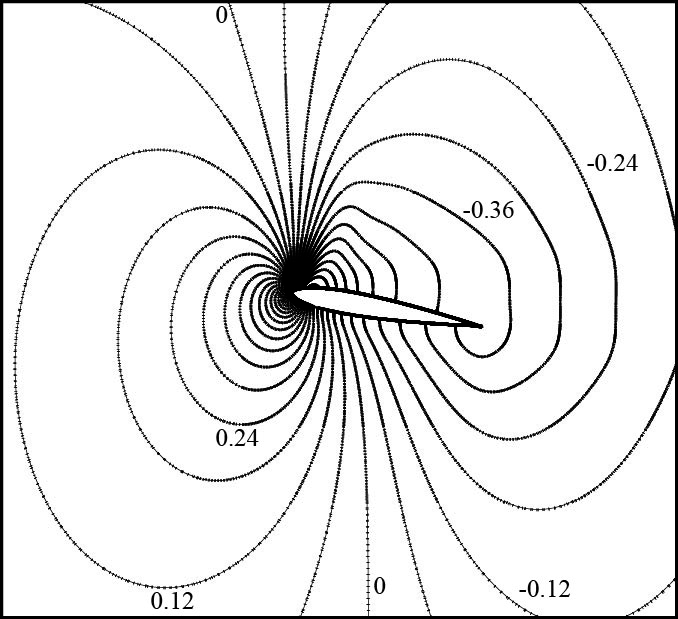}
}\hspace{0.05\textwidth}%
\subfigure[]{
\includegraphics[width=0.45\textwidth]{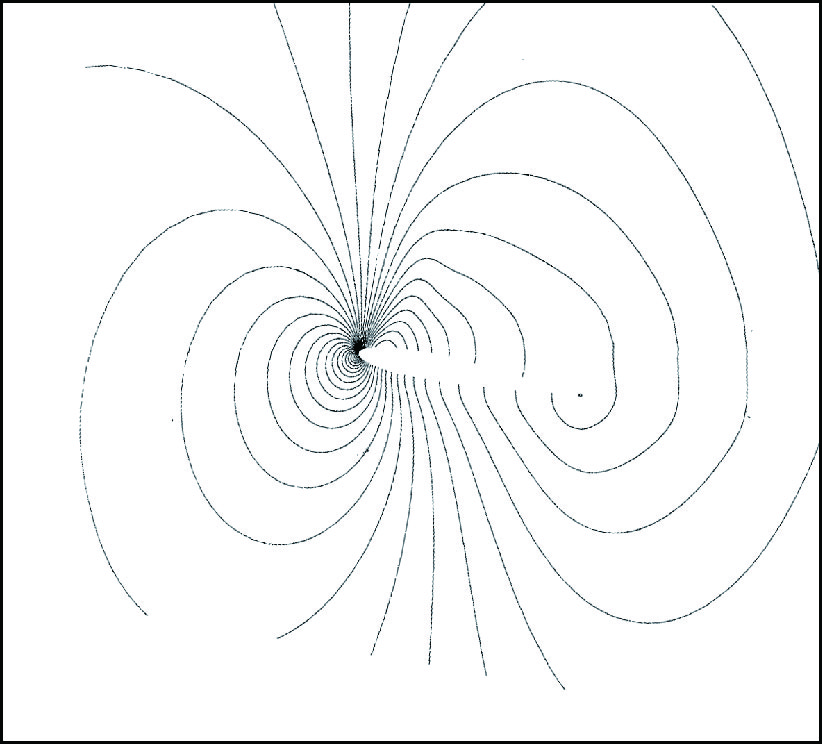}
}
\caption{\label{fig:test4_cpc} Pressure coefficient contours for flow around a NACA0012 airfoil at $\rm{Ma} = 0.8$, $\rm{Re} = 500$ and $\alpha  = 10^\circ $, (a) the present result (solid lines are based on the moving airfoil, dotted lines are based on the stationary airfoil, $\Delta {C_p} = 0.06$) and (b) the result of Hafez and Guo (Overflow)~\cite{hafez1999simulation}. Values of the isolines may be different between (a) and (b).}
\end{figure}

The Mach number contours and the pressure coefficient contours from the moving-airfoil simulation and the stationary-airfoil simulation are shown and compared with other numerical results in Fig.~\ref{fig:test4_ma} and Fig.~\ref{fig:test4_cpc}. Excellent agreement is acquired between the two sets of present results, and they seem also coincident with the numerical results of M{\"u}ller et al.~\cite{muller1987implicit} and Hafez and Guo (Overflow)~\cite{hafez1999simulation}.

\begin{figure}
\centering
\includegraphics[width=0.45\textwidth]{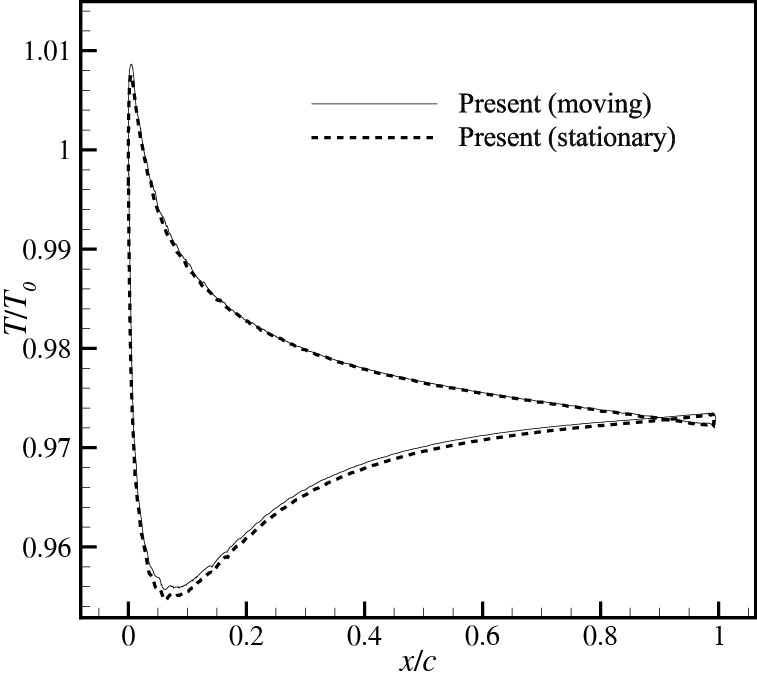}
\caption{\label{fig:test4_temper} Temperature distributions for flow around a NACA0012 airfoil at $\rm{Ma} = 0.8$, $\rm{Re} = 500$, $\rm{Pr}=0.71$ and $\alpha  = 10^\circ $. $T_0$ is the freestream total temperature. Solid lines are based on the moving airfoil, dashed lines are based on the stationary airfoil.}
\end{figure}

Due to the imposition of the adiabatic boundary condition, the temperature $T$ will vary along the surface of the airfoil. The distributions of $T$ based on the moving airfoil and the stationary airfoil are plotted in Fig.~\ref{fig:test4_temper}, where passible agreement is obtained between the two curves. It can be observed that at the leading edge the temperature $T$ is around or even higher than the free stream total temperature $T_0$, although it is lower than $T_0$ at most surface of the airfoil.

The distributions of pressure coefficient $C_p$ and skin friction coefficient $C_f$ on the surface of the airfoil are plotted and compared in Fig.~\ref{fig:test4_cpcf}. For $C_f$, the two present results and the result of M{\"u}ller et al.~\cite{muller1987implicit} almost overlap each other, showing perfect agreement. For $C_p$, the two present results and the result of Hafez and Guo (Overflow)~\cite{hafez1999simulation} agree well, while they deviate from the result of M{\"u}ller et al.~\cite{muller1987implicit} a little. This deviation may be partly due to that in \cite{muller1987implicit} the wall temperature of the airfoil is set to be equal to the freestream total temperature, i.e. the isothermal boundary condition is imposed.

\section{Conclusions}\label{sec:con}
In this paper, an IB method based on the ghost-cell approach is presented to simulate the viscous flow from incompressible to compressible with complex and moving boundary. The boundary condition is imposed by constructing ghost cells inside the solid body involved in the calculation stencil of the numerical solver. A ``local boundary determination'' thought is adopted to identify and construct the ghost cell, which allows different constructions for a ghost cell according to different boundary segments and eliminates the singularity problem of the ghost cell. For the moving boundary, a temporal extrapolation is used to calculate the variables on the fresh cell. Furthermore, the gas-kinetic BGK scheme is employed to solve the flow field. Some valuable discussions have been made, which is conductive to comprehending the physical mechanism of this numerical scheme. In addition, constructive modification has been made about the Prandtl number fixing of this scheme.

The test case of Taylor-Couette flow is conducted to analyze the spatial accuracy of the present method. We find that fluid cells with quite a part of area lying in the solid part of the convex boundary will have a very high error, which may be the evidence for the singularity problem of the ghost flow field. These singular fluid cells seem not to spoil the order of the local accuracy of the IB method and we exclude fluid cells adjoining at least one solid cell from the accuracy measure. It is testified that the method is at least second-order accurate in ${L_\infty }$ measure and also around second-order accurate in ${L_2 }$ measure. Moreover, the super-convergence phenomenon of the gas-kinetic BGK scheme has been observed and we find that a higher order reconstruction is necessary if we want to get a second-order accuracy about the temperature for this test case.

The supersonic flow over a stationary circular cylinder has been simulated. Although the vorticity distribution is somewhat oscillatory and the separating location is not very accurate due to the insufficient grid resolution, a lot of flow characteristics including the Mach number contours, pressure distribution and drag coefficient agree well with the published data. Then the test cases of incompressible flow around an oscillating cylinder and compressible flow around an airfoil are performed. The results agree very well with the previous experimental and numerical results. These test cases further validate the ability of the present method in simulating viscous incompressible/compressible flow with stationary/moving boundary.

\section*{Acknowledgements}
The Project financially supported by Natural Science Basic Research Plan in Shaanxi Province of China (Program No. 2015JM1002), National Natural Science Foundation of China (Grant No. 11472219), as well as National Pre-Research Foundation of China.

\clearpage


\bibliographystyle{yuan_immersed2}
\bibliography{yuan_immersed2}

\end{document}